\documentclass{imsart}

\RequirePackage[OT1]{fontenc}
\usepackage{amsthm,amsmath,natbib}
\RequirePackage[colorlinks,citecolor=blue,urlcolor=blue]{hyperref}
\usepackage{amssymb}
\usepackage{amsbsy}
\usepackage{epsfig}
\usepackage{color}
\usepackage{url}
\usepackage{color}
\usepackage{bm}
\usepackage{graphicx}
\usepackage{times}
\usepackage{rotating}
\usepackage{dsfont}

\renewcommand{\epsilon}{\varepsilon}
\renewcommand{\vec}[1]{\bm{#1}}
\newcommand{\mat}[1]{\MakeUppercase{\bm{#1}}}
\renewcommand{\hat}[1]{\widehat{#1}}

\renewcommand{\tilde}[1]{\widetilde{#1}}

\newcommand{\Cov}{\mathbb{C}}

\newcommand{\Var}{\mathbb{V}\text{ar}}
\newcommand{\N}{\mathcal{N}}

\newcommand{\diag}{\text{diag}}

\def\ind{\perp\!\!\!\perp}

\newcommand{\bs}{ {\boldsymbol s} }

\newcommand{\bX}{ {\boldsymbol X} }
\newcommand{\by}{ {\boldsymbol y} }

\newcommand{\bbeta}{ {\boldsymbol \beta} }

\newcommand{\bphi}{ {\boldsymbol \phi} }

\newcommand{\bOmega}{ {\boldsymbol \Omega} }

\newcommand{\bSigma}{ {\boldsymbol \Sigma} }

\newcommand{\given}{\,|\,}

\startlocaldefs
\numberwithin{equation}{section}
\theoremstyle{plain}

\endlocaldefs

\begin{document}

\begin{frontmatter}
\title{A Case Study Competition among Methods for Analyzing Large Spatial Data}
\runtitle{Case Study for Large Spatial Data}

\begin{aug}
\author{\fnms{Matthew J.} \snm{Heaton}\ead[label=e1]{mheaton@stat.byu.edu}},
\author{\fnms{Abhirup} \snm{Datta}\ead[label=e2]{abhidatta@jhu.edu}},
\author{\fnms{Andrew O.} \snm{Finley}\ead[label=e3]{finleya@msu.edu}},
\author{\fnms{Reinhard} \snm{Furrer}\ead[label=e4]{reinhard.furrer@math.uzh.ch}},
\author{\fnms{Rajarshi} \snm{Guhaniyogi}\ead[label=e5]{rguhaniy@ucsc.edu}},
\author{\fnms{Florian} \snm{Gerber}\ead[label=e6]{florian.gerber@math.uzh.ch}},
\author{\fnms{Robert B.} \snm{Gramacy}\ead[label=e7]{rbg@vt.edu}},
\author{\fnms{Joseph} \snm{Guinness}\ead[label=e14]{jsguinne@ncsu.edu}},
\author{\fnms{Dorit} \snm{Hammerling}\ead[label=e8]{dorith@ucar.edu}},
\author{\fnms{Matthias} \snm{Katzfuss}\ead[label=e9]{katzfuss@stat.tamu.edu}},
\author{\fnms{Finn} \snm{Lindgren}\ead[label=e10]{finn.lindgren@ed.ac.uk}},
\author{\fnms{Douglas} \snm{Nychka}\ead[label=e11]{nychka@ucar.edu}},
\author{\fnms{Furong} \snm{Sun}\ead[label=e12]{furongs@vt.edu}},
\and
\author{\fnms{Andrew} \snm{Zammit-Mangion}
\ead[label=e13]{azm@uow.edu.au}}
\runauthor{M. J. Heaton et al.}


\address{Matthew~J.~Heaton is Assistant Professor, Department of Statistics, Brigham Young University 223 TMCB, Provo, UT 84602 \printead{e1}.}
\address{Abhirup Datta is Assistant Professor, Department of Biostatistics, Johns Hopkins University, 615 N. Wolfe Street, Baltimore, MD 21205 \printead{e2}.}
\address{Andrew Finley is Associate Professor, Department of Forestry and Geography, Michigan State University, 126 Natural Resources Building, Each Lansing, MI 48824 \printead{e3}.}
\address{Reinhard Furrer is Professor, Department of Mathematics and Department of Computational Science, University of Zurich, Switzerland \printead{e4}.}
\address{Rajarshi Guhaniyogi is Assistant Professor, Department of Applied Mathematics \& Statistics, University of California Santa Cruz, 1156 High Street, SOE2, Santa Cruz, CA 95064 \printead{e5}.}
\address{Florian Gerber is postdoctoral research fellow, Department of Mathematics, University of Zurich, Switzerland \printead{e6}.}
\address{Robert B. Gramacy is Professor, Department of Statistics, Virginia Tech, Department of Statistics (MC0439), Hutcheson Hall, 250 Drillfield Drive, Blacksburg, VA 24061 \printead{e7}.}
\address{Joseph Guinness is Assistant Professor, Department of Statistics, North Carolina State University, 2311 Stinson Drive, Campus Box 8203 Raleigh, NC 27695 \printead{e14}.}
\address{Dorit Hammerling is Section Leader of Statistics and Data Science, Institute for Mathematics Applied to Geosciences, National Center for Atmospheric Research, Boulder, CO 80305 \printead{e8}.}
\address{Matthias Katzfuss is Assistant Professor, Department of Statistics, Texas A\&M University, College Station, TX 77843 \printead{e9}.}
\address{Finn Lindgren is Chair of Statistics, School of Mathematics, University of Edinburgh, Edinburgh EH9 3FD, United Kingdom \printead{e10}.}
\address{Douglas Nychka is Director, Institute for Mathematics Applied to Geosciences, National Center for Atmospheric Research, Boulder, CO 80305 \printead{e11}.}
\address{Furong Sun is graduate student, Department of Statistics, Virginia Tech, Hutcheson Hall, 250 Drillfield Drive, Blacksburg, VA 24061 \printead{e12}.}
\address{Andrew Zammit-Mangion is Senior Lecturer, School of Mathematics and Applied Statistics, University of Wollongong, NSW 2522, Australia \printead{e13}.}
\end{aug}

\begin{abstract}
The Gaussian process is an indispensable tool for spatial data analysts.  The onset of the ``big data'' era, however, has lead to the traditional Gaussian process being computationally infeasible for modern spatial data.  As such, various alternatives to the full Gaussian process that are more amenable to handling big spatial data have been proposed.  These modern methods often exploit low rank structures and/or multi-core and multi-threaded computing environments to facilitate computation.  This study provides, first, an introductory overview of several methods for analyzing large spatial data.  Second, this study describes the results of a predictive competition among the described methods as implemented by different groups with strong expertise in the methodology.  Specifically, each research group was provided with two training datasets (one simulated and one observed) along with a set of prediction locations.  Each group then wrote their own implementation of their method to produce predictions at the given location and each which was subsequently run on a common computing environment.  The methods were then compared in terms of various predictive diagnostics.  Supplementary materials regarding implementation details of the methods and code are available for this article online.
\end{abstract}

\begin{keyword}
\kwd{Big data}
\kwd{Gaussian process}
\kwd{Parallel computing}
\kwd{Low rank approximation}
\end{keyword}

\end{frontmatter}

\section{Introduction}\label{intro}

For decades, the Gaussian process (GP) has been the primary tool used for the analysis of geostatistical (point-referenced) spatial data \citep{schabenberger2004statistical,cressie2015statistics,cressiewikle2015statistics,banerjee2014hierarchical}.   A spatial process $Y(\vec{s})$ for $\vec{s} \in \mathcal{D} \subset \mathbb{R}^2$ is said to follow a GP if any realization $\vec{Y} = (Y(\vec{s}_1),\dots,Y(\vec{s}_N))'$ at the finite number of locations $\vec{s}_1,\dots,\vec{s}_N$ follows an $N$-variate Gaussian distribution.  More specifically, let $\mu(\vec{s}): \mathcal{D} \rightarrow \mathbb{R}$ denote a mean function returning the mean at location $\vec{s}$ (typically assumed to be linear in covariates $\vec{X}(\vec{s}) = (1,X_1(\vec{s}),\dots,X_P(\vec{s}))'$) and $\mathbb{C}(\vec{s}_1,\vec{s}_2): \mathcal{D}^2 \rightarrow \mathbb{R}^+$ denote a positive definite covariance function. Then, if $Y(\vec{s})$ follows a spatial Gaussian process, $\vec{Y}$ has the density function,
\begin{align}
f_{\vec{Y}}(\vec{y}) &= \left(\frac{1}{\sqrt{2\pi}}\right)^{N} |\vec{\Sigma}|^{-1/2} \exp\left\{-\frac{1}{2}(\vec{y}-\vec{\mu})'\vec{\Sigma}^{-1}(\vec{y}-\vec{\mu})\right\}
\label{GaussPDF}
\end{align}
where $\vec{\mu} = (\mu(\vec{s}_1),\dots,\mu(\vec{s}_N))'$ is the mean vector and $\vec{\Sigma} = \{\mathbb{C}(\vec{s}_i,\vec{s}_j)\}_{ij}$ is the $N\times N$ covariance matrix  governed by $\mathbb{C}(\vec{s}_i,\vec{s}_j)$ (e.g.\ the Mat\'{e}rn covariance function).  From this definition, the appealing properties of the Gaussian distribution (e.g.\ Gaussian marginal and conditional distributions) have rendered the GP an indispensable tool for any spatial data analyst to perform such tasks as kriging (spatial prediction) and proper uncertainty quantification.

With the modern onset of larger and larger spatial datasets, however, the use of Gaussian processes for scientific discovery has been hindered by computational intractability.  Specifically, evaluating the density in \eqref{GaussPDF} requires $\mathcal{O}(N^3)$ operations and $\mathcal{O}(N^2)$ memory which can quickly overwhelm computing systems when $N$ is only moderately large.  Early solutions to this problem included factoring \eqref{GaussPDF} into a series of conditional distributions \citep{vecchia1988estimation,stein2004approximating}, the use of pseudo-likelihoods \citep{varin2011overview,eidsvik2014estimation}, modeling in the spectral domain \citep{fuentes2007approximate} or using tapered covariance functions \citep{furrer2006covariance,kaufman2008covariance,stein2013statistical}.  Beginning in the late 2000's, several approaches based on low rank approximations to Gaussian processes were developed (or became popular) including discrete process convolutions \citep{higdon2002space,lemos2009spatio}, fixed rank kriging \citep{cressie2008fixed,kang2011bayesian,katzfuss2011spatio}, predictive processes \citep{banerjee2008gaussian,finley2009improving}, lattice kriging \citep{nychka2015multiresolution} and stochastic partial differential equations \citep{lindgren2011explicit}.  \citet{sun2012geostatistics} and \citet{bradley2016comparison} provide exceptional reviews of these methods and demonstrate their effectiveness for modeling spatial data.  

After several years of their use, however, scientists have started to observe shortcomings in many of the above methods for approximating GPs such as the propensity to oversmooth the data \citep{simpson2012order,stein2014limitations} and even, for some of these methods, an upper limit on the size of the dataset that can be modeled.  Hence, recent scientific research in this area has focused on the efficient use of modern computing platforms and the development of methods that are parallelizable.  For example, \citet{paciorek2013parallelizing} show how \eqref{GaussPDF} can be calculated using parallel computing while \citet{katzfuss2014parallel} and \citet{Katzfuss2015} develop a basis-function approach that lends itself to distributed computing. Alternatively, \citet{barbian2017spatial} and \citet{guhaniyogi2018meta} propose dividing the data into a large number of subsets, draw inference on the subsets in parallel and then combining the inferences. \citet{datta2016hierarchical,datta2016nonseparable} build upon \citet{vecchia1988estimation} by developing novel approaches to factoring \eqref{GaussPDF} as a series of conditional distributions based only on nearest neighbors.  

Given the plethora of choices to analyze large spatially correlated data, for this paper, we seek to not only provide an overview of modern methods to analyze massive spatial datasets, but also lightly compare the methods in a unique way.  Specifically, this research implements the common task framework of \citet{wikle2017common} by describing the outcome of a friendly case study competition between various research groups across the globe who each implemented their own method to analyze the same spatial datasets (see the list of participating groups in Table \ref{groups}).  That is, several research groups were provided with two spatial datasets (one simulated and one real) with a portion of each dataset removed to validate predictions (research groups were not provided with the removed portion so that this study is ``blinded'').  The simulated data represents a scenario where the Gaussian process assumption is valid (i.e., a correctly specified model), whereas the real dataset is a scenario when the model is potentially misspecified due to inherent non-stationarity or non-Gaussian errors.  Each group then implemented their unique method and provided a prediction (and prediction interval or standard error) of the spatial process at the held out locations.  The predictions were compared by a third party and are summarized herein.  

The case study competition described herein is unique and novel in that, typically, comparisons/reviews of various methods is done by a single research group implementing each method.  However, single research groups may be more or less acquainted with some methods leading to a possibly unfair comparison with those methods they are less familiar with.  In contrast, for the comparison/competition here, each method was implemented by a research group with strong expertise in the method and who is well-versed in any possible intricacies associated with its use.  Hence, in terms of scientific contributions, this paper (i) serves as a valuable review, (ii) discusses a unique case study comparison of spatial methods for large datasets, (iii) provides code to implement each method to practitioners (see supplementary materials) and (iv) establishes a framework for future studies to follow when comparing various analytical methods.

The remainder of this paper is organized as follows.  Section \ref{methods} gives a brief background on each method.  Section \ref{competition} provides the setting for the comparison along with background on the datasets.    Section \ref{results} then summarizes the results of the comparison in terms of predictive accuracy, uncertainty quantification and computation time.  Section \ref{conc} draws conclusions from this study and highlights future research areas for the analysis of massive spatial data.

\section{Overview of Methods for Analyzing Large Spatial Data}\label{methods}

\subsection{Fixed Rank Kriging}
Fixed Rank Kriging \citep[FRK,][]{Cressie_2006, cressie2008fixed} is built around the concept of a \emph{spatial random effects} (SRE) model. In FRK, one models the process $\tilde{Y}(\vec{s}), \vec{s} \in D,$ as 
\begin{align}
\tilde{Y}(\vec{s}) = \mu(\vec{s}) + w(\vec{s}) + \xi(\vec{s}),\quad \vec{s} \in D,
\label{SREmodel}
\end{align}
where $\mu(\vec{s})$ is the mean function that is itself modeled as a linear combination of known covariates, $w(\vec{s})$ is the SRE model, and $\xi(\vec{s})$ is a fine-scale process, modeled to be spatially uncorrelated with variance $\sigma^2_\xi$.  The process $\xi(\vec{s})$ in \eqref{YFRK} is designed to soak up variability in $\tilde{Y}(\vec{s})$ not accounted for by $w(\vec{s})$.  

The primary assumption of FRK is that the spatial process $w(\cdot)$ can be decomposed into a linear combination of $K$ basis functions $\vec{h}(\vec{s}) = (h_1(\vec{s}),\dots,h_K(\vec{s}))', \vec{s} \in D,$ and $K$ basis function coefficients $\vec{\theta} = (\theta_1,\dots,\theta_K)'$ such that,
\begin{align}
w(\vec{s}) = \sum_{k=1}^K h_k(\vec{s})\theta_k, \quad \vec{s} \in \mathcal{D}.
\label{basis}
\end{align}
The use of $K$ basis functions ensures that all estimation and prediction equations only contain inverses of matrices of size $K \times K$, where $K \ll N$.  In practice, the set $\{h_k(\vec{\cdot})\}$ in \eqref{basis} is comprised of functions at $R$ different resolutions such that \eqref{basis} can also be written as
\begin{align}
w(\vec{s}) = \sum_{r=1}^R\sum_{k=1}^{K_r} h_{rk}(\vec{s})\theta_{rk},\quad \vec{s} \in D,
\label{multires}
\end{align}
where $h_{rk}(\vec{s})$ is the $k^{th}$ spatial basis function at the $r^{th}$ resolution with associated coefficient $\theta_{rk}$, and $K_r$ is the number of basis functions at the $r^{th}$ resolution, such that $K=\sum_{r=1}^R K_r$ is the total number of basis functions used.  For this research, we used $R=3$ resolutions of bisquare basis functions following \citet{cressie2008fixed}.

The coefficients $\vec{\theta} = \{\theta_{rk}\}$ have $\Var(\vec{\theta}) = \vec{S}(\vec{\phi})$ with covariance parameters $\vec{\phi}$ that need to be estimated.  In this work, $\vec{S}(\vec{\phi})$ is a block-diagonal matrix composed from $R$ dense matrices, where the $r^{th}$ block has $i,j$th element $\exp(-d_r(i,j)/\phi_r)$ and $d_r(i,j)$ is the distance between the centroids of the $i^{th}$ and $j^{th}$ basis function at the $r^{th}$ resolution, and $\vec{\phi} = (\phi_1,\dots,\phi_R)'$ are the spatial correlation parameters of the exponential correlation function. Note that $\vec{S}(\vec{\phi})$ can also be unstructured in which case $K(K+1)/2$ parameters need to be estimated, however this case is not considered here.

There are several variants of FRK. In this work, we use the implementation by \cite{zammit2017frk} which comes in the form of the \texttt{R} package \texttt{FRK}, available from the Comprehensive \texttt{R} Archive Network (CRAN). In this paper we utlize v0.1.6 of that package.  In \texttt{FRK} the model for $\tilde{Y}(\vec{s}), \vec{s} \in D,$ is composed as in \eqref{SREmodel}.  \texttt{FRK} further assumes that recorded observations $Y(\vec{s}_i)$ are noisy readings of $\tilde{Y}(\vec{s}_i), i = 1,\dots, N,$ such that
\begin{align}
Y(\vec{s}_i) = \tilde{Y}(\vec{s}_i) + \epsilon(\vec{s}_i), \quad i = 1,\dots, N,
\label{YFRK}
\end{align}
where for $i = 1,\dots,N$, $\epsilon(\vec{s}_i)$ denotes independent and identically normally distributed measurement error with mean 0 and known measurement error variance $\sigma^2_\epsilon$. More details on the implementation of FRK for this study are included in the supplementary materials.

\subsection{LatticeKrig}
LatticeKrig \citep[LK,][]{nychka2015multiresolution} uses nearly the same setup as is employed by FRK.  Specifically, LK assumes the model \eqref{SREmodel} and \eqref{YFRK} but omits the fine-scale process $\xi(\cdot)$.  Further, for $w(\vec{s})$, LK follows the multiresolution approach in \eqref{multires}, but LK uses a different structure and constraints than FRK.  First, the marginal variance of each resolution $\vec{h}_{r}'(\vec{s})\vec{\theta}_r$ where $\vec{h}_r'(\vec{s}) = (h_{r1}(\vec{s}),\dots,h_{rK_r}(\vec{s}))'$ are the basis functions of the $r^{th}$ resolution with coefficients $\vec{\theta}_{r} = (\theta_{r1},\dots,\theta_{rK_r})'$ is constrained to be $\sigma^2_w\alpha_r$ where $\sigma^2_w,\alpha_r>0$ and $\sum_{r=1}^R\alpha_r = 1$.  To further reduce the number of parameters, LK sets $\alpha_r \sim r^{-\nu}$ where $\nu$ is a single free parameter.

LatticeKrig obtains multiresolution radial basis functions by translating and scaling a radial function in the following manner.  Let $\vec{u}_{rk}$ for $r=1,\dots,R$ and $k=1,\dots,K_r$ denote a regular grid of $K_r$ points on $\mathcal{D}$ corresponding to resolution $r$.  For this article, LK defines
\begin{align}
h_{rk}(\vec{s}) = \psi(\|\vec{s}-\vec{u}_{rk}\|/\theta_r)
\end{align}
where the distance is taken to  be Euclidean because the spatial region in this case is of small geographic extent and  $\theta_r = 2^{-r}$.  Further, LK defines
\begin{align}
\psi(d) \propto \begin{cases}\frac{1}{3}(1-d)^6 ( 35d^2 + 18d+3) & \text{ if } d \leq 1\\
0 & \text{ otherwise.}
\end{cases}
\label{Wendland}
\end{align}
which are Wendland polynomials and are positive definite (an attractive property when the basis is used for interpolation).  Finally, the basis functions in \eqref{Wendland} are normalized at each resolution so that the process marginal variance at all $\vec{s}$ is $\sigma^2_w \alpha_r$. This  reduces edge effects and makes for a better approximation to a stationary covariance function.

LatticeKrig assumes the coefficients at each resolution $\vec{\theta}_{r} = (\theta_{r1},\dots,\theta_{rK_r})'$ are independent (similar to the block diagonal structure used in FRK) and follow a multivariate normal distribution with covariance $\vec{Q}_r^{-1}$ parameterized by a single parameter $\phi_r$.  Because the locations $\{\vec{u}_{rk}\}_{k=1}^{K_r}$ are prescribed to be a regular grid, LK uses a spatial autoregression/Markov random field \citep[see][Section 4.4]{banerjee2014hierarchical} structure for $\vec{Q}_r^{-1}$ leading to sparsity and computational tractability.  Furthermore, because $\vec{Q}_r$ is sparse, LK can set $K$ to be very large (as in this competition greater than $N$) without much additional computational cost.  The supplementary material to this article contains additional information about the implementation of LatticeKrig used in this case study.

\subsection{Predictive Processes}
For the predictive process (PP) approach, let $\vec{s}^\star_1,\dots,\vec{s}^\star_K$ denote a set of ``knot'' locations well dispersed over the spatial domain $\mathcal{D}$.  Assume that the SREs ($w(\vec{s})$) in \eqref{SREmodel} follow a mean zero Gaussian process with covariance function $\Cov(\vec{s},\vec{s}') = \sigma^2_w \rho(\vec{s},\vec{s}')$ where $\rho(\cdot,\cdot)$ is a postive definite correlation function. Under this Gaussian process assumption, the SREs $\vec{w}^\star = (w(\vec{s}^\star_1),\dots,w(\vec{s}^\star_K))' \sim \N(0,\mat{\Sigma}_{w^\star})$ where $\mat{\Sigma}_{w^\star}$ is a $K\times K$ covariance matrix with $ij^{th}$ element $\Cov(\vec{s}^\star_i,\vec{s}_j^\star)$.  The PP approach exploits the Gaussian process assumption for the SREs and replaces $w(\vec{s})$ in \eqref{SREmodel} with
\begin{align}
\tilde{w}(\vec{s}) = \Cov'(\vec{s},\vec{s}^\star)\mat{\Sigma}_{w^\star}^{-1}\vec{w}^\star
\label{PP}
\end{align}
where $\Cov(\vec{s},\vec{s}^\star) = (\Cov(\vec{s},\vec{s}_1^\star),\dots,\Cov(\vec{s},\vec{s}_K^\star))'$.  Note that \eqref{PP} can be equivalently written as the basis function expression given above in \eqref{basis} where the basis functions are $\Cov(\vec{s},\vec{s}^\star)\mat{\Sigma}_{w^\star}^{-1}$ and $\vec{w}^\star$ effectively plays the role of the basis coefficients.

\citet{finley2009improving} noted that the basis function expansion in \eqref{PP} systematically underestimates the marginal variance $\sigma^2_w$ from the original process.  That is, $\Var(\tilde{w}(\vec{s})) = \Cov'(\vec{s},\vec{s}^\star)\mat{\Sigma}_{w^\star}^{-1}\Cov'(\vec{s},\vec{s}^\star) $ $\leq \sigma^2_w$.  To counterbalance this underestimation of the variance, \citet{finley2009improving} use the structure in \eqref{YFRK},
\begin{align}
Y(\vec{s}) = \mu(\vec{s}) + \tilde{w}(\vec{s}) + \xi(\vec{s}) + \epsilon(\vec{s})
\label{modifiedPP}
\end{align}
where $\xi(\vec{s})$ are spatially independent with distribution $\N(0,\sigma^2_w-\Cov'(\vec{s},\vec{s}^\star)\mat{\Sigma}_{w^\star}^{-1}\Cov'(\vec{s},\vec{s}^\star))$ such that $\Var(\tilde{w}(\vec{s}) + \xi(\vec{s})) = \sigma^2_w$ as in the original parent process.

As with FRK and LatticeKrig, the associated likelihood under \eqref{modifiedPP} only requires calculating the inverse and determinant of a dense $K\times K$ matrix and diagonal $N\times N$ matrices which results in massive computational savings when $K \ll N$ and $K$ is small.  However, one advertised advantage of using the PP approach as opposed to FRK or LatticeKrig is that the PP basis functions are completely determined by the choice of covariance function $\Cov(\cdot,\cdot)$.  Hence, the PP approach is unaltered even when considering modeling complexities such as anisotropy, non-stationarity or even multivariate processes.  At the same time, however, when $\Cov(\cdot,\cdot)$ is governed by unknown parameters (which is nearly always the case) the PP basis functions need to be calculated iteratively rather than once as in FRK or LatticeKrig which will subsequently increase computation time.

\subsection{Spatial Partitioning}
Let the spatial domain $\mathcal{D} = \bigcup_{d=1}^D \mathcal{D}_d$ where $\mathcal{D}_1,\dots,\mathcal{D}_D$ are subregions that form a partition (i.e.\ $\mathcal{D}_{d_1} \bigcap \mathcal{D}_{d_2} = \emptyset$ for all $d_1 \neq d_2$).  The modeling approach based on spatial partitioning is to assume \textit{conditional} dependence between observations within a subregion and \textit{conditional} independence between observations across subregions.  More specifically, if $\vec{Y}_d = \{Y(\vec{s}_i): \vec{s}_i \in \mathcal{D}_d\}$ where $d=1,\dots,D$, then 
\begin{align}
\vec{Y}_d \overset{ind}{\sim} \mathcal{N}\left(\mat{X}_d\vec{\beta}+\mat{H}_d\vec{\theta},\mat{\Sigma}(\vec{\phi}_d)\right)
\label{partModel}
\end{align}
where $\mat{X}_d$ is a design matrix containing covariates associated with $\vec{Y}_d$, $\mat{H}_d$ is a matrix of spatial basis functions (such as those used in predictive processes, fixed rank kriging or lattice kriging mentioned above) and $\mat{\Sigma}(\vec{\phi}_d)$ is the covariance matrix for subregion $d$ governed by covariance parameters $\vec{\phi}_d$ (e.g.\ decay and smoothness parameters).  Notice that, in \eqref{partModel} each subregion shares common $\vec{\beta}$ and $\vec{\theta}$ parameters which allows smoothing across subregions (hence, $Y_{d_1} \ind Y_{d_2}$ for $d_1 \neq d_2$ conditional on the parameters $\vec{\beta}$ and $\vec{\theta}$).  Further, the assumption of independence across subregions allows the likelihood for $\vec{\beta}, \vec{\theta}$ and $\vec{\phi}_d$ is to be computed in parallel thereby facilitating computation.

By way of distinction, this approach is inherently different from the ``divide and conquer'' approach \citep{liang2013resampling,barbian2017spatial}.  In the divide and conquer approach, the full dataset is subsampled, the model is fit to each subset and the results across subsamples are pooled.  In contrast, the spatial partition approach uses all the data simultaneously in obtaining estimates, but the independence across regions facilitates computation.

The key to implementing the spatial partitioning approach is the choice of partition and the literature is replete with various options.  \textit{A priori} methods to define the spatial partitioning include partitioning the region into equal areas \citep{sang2011covariance},  partitioning based on centroid clustering \citep{knorr2000bayesian,kim2005analyzing} and hierarchical clustering based on spatial gradients \citep{anderson2014identifying,heaton2017nonstationary}.  Alternatively, model-based approaches to spatial partitioning include treed regression \citep{konomi2014adaptive} and mixture modeling \citep{neelon2014multivariate} but these approaches typically require more computation.  For this analysis, a couple of different partitioning schemes were considered, but each scheme resulted in approximately equivalent model fit to the training data.  Hence, based on the results from the training data, for the competition below we used an equal area partition of approximately $6000$ observations per subregion.

\subsection{Covariance Tapering}
The idea of covariance tapering is based on the fact that many entries in the covariance matrix $\vec{\Sigma}$ in~\eqref{GaussPDF} are close to zero and associated location pairs could be considered as essentially independent. Covariance tapering multiplies the covariance function $\mathbb{C}(\vec{s}_i,\vec{s}_j)$ with a compactly supported covariance function, resulting in another positive definite covariance function but with compact support. From a theoretical perspective, covariance tapering (in the framework of infill-asymptotics) is using the concept of Gaussian equivalent measures and mis-specified covariance functions (see, e.g., \citealp{Stei:99} and references therein). Subsequently, \cite{furrer2006covariance} have assumed a second-order stationary and isotropic Mat\'ern covariance to show asymptotic optimality for prediction under tapering.  This idea has been extended to different covariance structures \citep{stein2013statistical}, non-Gaussian response \citep{Hira:Yaji:13} and
multivariate and/or spatio-temporal setting \citep{Furr:Bach:Du:16}.


From a computational aspect, the compact support of the resulting covariance function provides the computational savings needed by employing sparse matrix algorithms to efficiently solve systems of linear equations.
More precisely, to evaluate density~\eqref{GaussPDF}, a Cholesky factorization for $\mat{\Sigma}$ is performed followed by two solves of triangular systems. For typical spatial data settings, the solve algorithm is effectively linear in the number of observations. 

For parameter estimation in the likelihood framework, one- and two-taper approaches exist \citep[see][for relevant literature]{kaufman2008covariance,Du:etal:09,Wang:Loh:11,Bevi:etal:16}. To distinguish the two approaches, notice that the likelihood in \eqref{GaussPDF} can be rewritten as
\begin{align}
f_{\vec{Y}}(\vec{y}) &= \left(\frac{1}{\sqrt{2\pi}}\right)^{N} |\vec{\Sigma}|^{-1/2} \text{etr}\left\{-\frac{1}{2}(\vec{y}-\vec{\mu})(\vec{y}-\vec{\mu})'\mat{\Sigma}^{-1}\right\}
\label{GaussPDF2}
\end{align}
where $\text{etr}(\mat{A}) = \exp(\text{trace}(A))$.  In the one-taper
setting, only the covariance is tapered such that $\mat{\Sigma}$ in \eqref{GaussPDF2} is replaced by $\mat{\Sigma}\odot\mat{T}$ where ``$\odot$'' denotes the Hadamard product and $\mat{T}$ is the $N\times N$ tapering matrix.  In the two-tapered approach both the covariance and empirical covariance are affected such that not only is $\mat{\Sigma}$ replaced by $\mat{\Sigma}\odot\mat{T}$ but $(\vec{y}-\vec{\mu})(\vec{y}-\vec{\mu})'$ is replaced by $(\vec{y}-\vec{\mu})(\vec{y}-\vec{\mu})' \odot \mat{T}$. The
one-taper equation results in biased estimates of model parameters while the two-taper approach is based on estimating equations (and is, therefore, unbiased) but comes at the price of a severe loss of  computational efficiency. If the one-taper biased estimates of model parameters are used for prediction, the biases may result in some loss of predictive accuracy \citep{Furr:Bach:Du:16}. 


Although tapering can be adapted to better take into account uneven densities of locations and complex anisotropies, we use a simple straight-forward approach for this competition. The implementation here relies almost exclusively on the \texttt{R} package {\tt spam} \citep{Furr:Sain:10,spam}.  Alternatively to likelihood approaches and in view of computational costs, we have minimized the squared difference between an empirical covariance and parameterized covariance function. The gridded structure of the data is exploited and the empirical covariance is estimated for a specific set of locations only; and thus is close to classical variogram estimation and fitting \citep{cressie2015statistics}. 

\subsection{Multiresolution Approximations}
The multi-resolution approximation (MRA) can be viewed as a combination of several previously described approaches. 
Similar to FRK or LatticeKrig, the MRA expresses the spatial process of interest $w(\vec{s})$ in \eqref{SREmodel} as a weighted sum of \textit{compactly} supported basis functions at different resolutions as in \eqref{multires}. In contrast to FRK or LatticeKrig, the MRA basis functions and the prior distribution of the corresponding weights are chosen using the predictive-process approach to automatically adapt to any given covariance function $\Cov(\cdot)$, and so the MRA can adjust flexibly to a desired spatial smoothness and dependence structure.  Scalability of the MRA is ensured in that for increasing resolution, the number of basis functions increases while the support of each function (i.e.,\ the part of the spatial domain in which it is nonzero) decreases. Decreasing support (and increasing sparsity of the covariance matrices of the corresponding weights) is achieved either by increasingly severe tapering of the covariance function 
(MRA-taper; \citealt{KatzfussGong2017}) or by recursively partitioning the spatial domain \citep[MRA-block;][]{Katzfuss2015}.  This can lead to (nearly) exact approximations with quasilinear computational complexity.

\begin{figure}[tb]
	\centering
	\includegraphics[width =.6\linewidth]{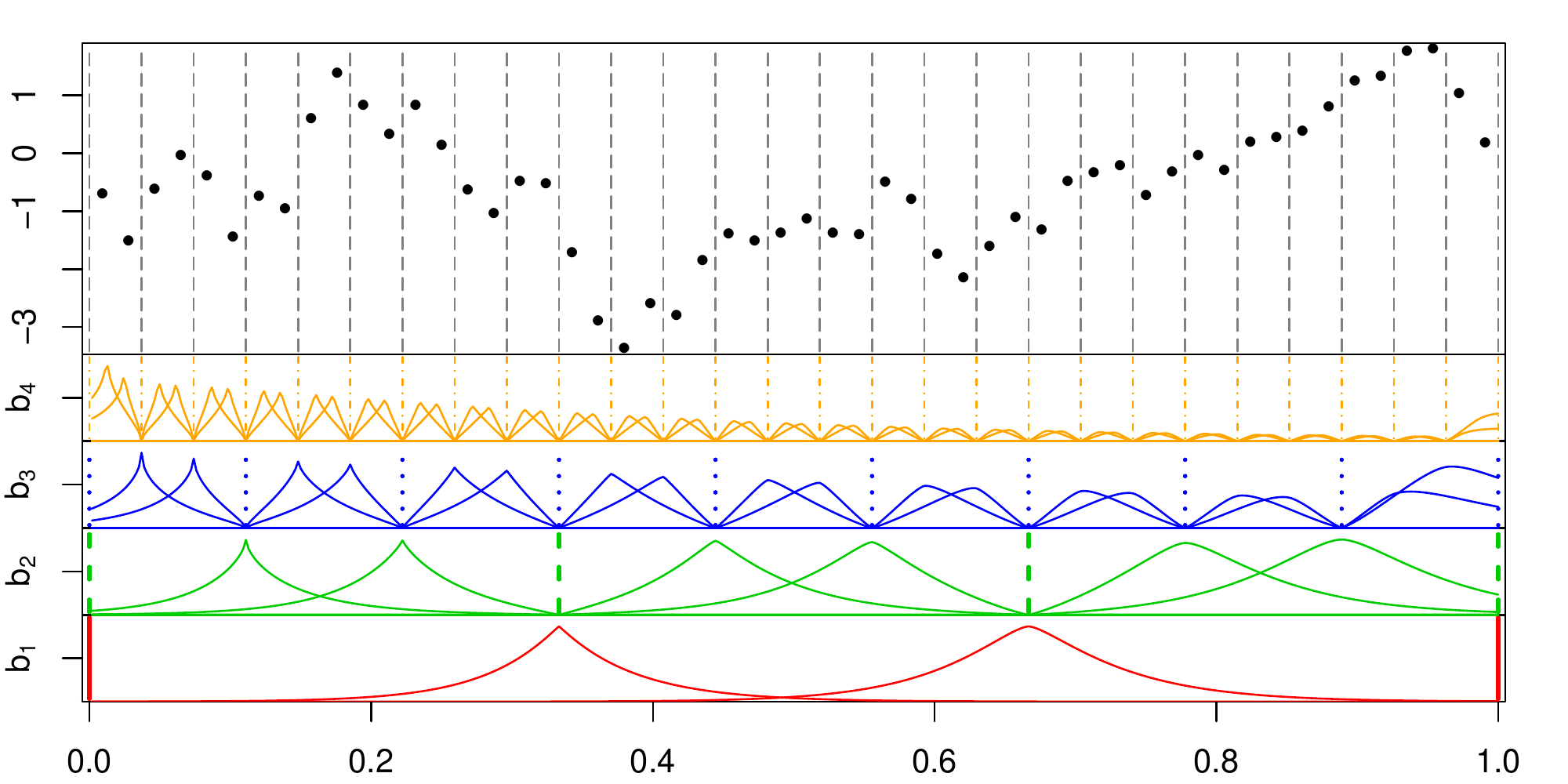}
	\caption{A toy example of simulated observations (black dots) with a covariance function $\Cov$ with increasing smoothness on a one-dimensional spatial domain $\mathcal{D} = [0,1]$, together with a multi-resolution approximation (MRA) with $M=4$ resolutions with 3 subregions per region (vertical lines) and $r_0=2$ basis functions per region. The basis functions and their weights (symbolized by the height of the functions) adjust to the changing smoothness, here increasing from left to right.}
	\label{fig:toymradata}
\end{figure}

While the MRA-taper has some attractive smoothness properties, we focus here on the MRA-block which is based on a recursive partitioning of the domain $\mathcal{D}$ into smaller and smaller subregions up to some level $M$. Within each (sub-)region at each resolution, there is a small number, say $r_0$, of basis functions.
The resulting approximation of the process (including its variance and smoothness) in each region at resolution $M$ is exact. In addition, it is feasible to compute and store the joint posterior covariance matrix (i.e., not just its inverse as with related approaches) for a large number of prediction locations as a product of two sparse matrices.
Figure \ref{fig:toymradata} illustrates the MRA basis functions in a toy example.

The MRA-block is designed to take full advantage of high-performance computing systems, in that inference is well suited for massively distributed computing, with limited communication overhead. The computational task is split into small parts by assigning a computational node to each region of the recursive partitioning. The nodes then deal in parallel with the basis functions corresponding to their assigned regions leading to a polylogarithmic computational complexity.  For this project, we use $M=9$ levels, partition each domain in 2 parts and set the number of basis function in each partition to $r_0=64$. 

\subsection{Nearest Neighbor Processes}\label{NNGP}
The nearest neighbor Gaussian process (NNGP) developed in \citet{datta2016hierarchical} and \citet{datta2016cholesky} is defined from the conditional specification of the joint distribution of the SREs in \eqref{SREmodel}.  Let $w(\vec{s})$ in \eqref{SREmodel} follow a mean zero Gaussian process with $\Cov(\vec{s},\vec{s}') = \sigma^2_w\rho(\vec{s},\vec{s}')$ where $\rho(\cdot)$ is a positive definite correlation function.  Factoring the joint distribution of $w(\vec{s}_1),\dots,w(\vec{s}_N)$ into a series of conditional distributions yields that $w(\vec{s}_1) = 0+\eta(\vec{s}_1)$ and 
\begin{align}\label{eq:nngpin}
w(\vec{s}_i) \mid \vec{w}_{1:(i-1)} = \Cov'(\vec{s}_1,\vec{s}_{1:(i-1)})\mat{\Sigma}_{1:(i-1)}^{-1}\vec{w}_{1:(i-1)} + \eta(\vec{s}_i)
\end{align} 
where $\vec{w}_{1:(i-1)} = (w(\vec{s}_1),\dots,w(\vec{s}_{i-1}))'$, $\Cov(\vec{s}_1,\vec{s}_{1:(i-1)}) = (\Cov(\vec{s}_i,\vec{s}_1),\dots,\Cov(\vec{s}_i,\vec{s}_{i-1})'$, $\mat{\Sigma}_{1:(i-1)} = \Var(\vec{w}_{1:(i-1)})$ and $\eta$'s are independent, mean zero, normally distributed random variables. More compactly, \eqref{eq:nngpin} is equivalent to $\vec w = \vec A \vec w  + \vec \eta$ where $\vec A = (a_{ij})$ is a lower triangular matrix with zeroes along the diagonal and $\vec \eta = (\eta(\vec{s}_1),\dots,\eta(\vec{s}_n))' \sim N(0, \vec D)$ with diagonal entries $\Cov(\vec{s}_i,\vec{s}_i) - \Cov'(\vec{s}_1,\vec{s}_{1:(i-1)})\mat{\Sigma}_{1:(i-1)}^{-1}\Cov(\vec{s}_1,\vec{s}_{1:(i-1)})$. This effectuates a joint distribution $\vec w \sim N(0, \vec{\Sigma})$ where $\vec \Sigma^{-1} = (\vec I - \vec A)'\vec D ^ {-1} (\vec I - \vec A)$. 
Furthermore, when predicting for any $\vec{s} \notin \{\vec{s}_1,\dots,\vec{s}_N\}$, one can define 
\begin{equation}\label{eq:nngpout}
w(\vec{s})\mid \vec{w}_{1:N} = \vec{a}'(s)\vec w_{1:N} + \eta(\vec{s})
\end{equation}
similar to \eqref{eq:nngpin}. 

A sparse formulation of $\vec A$ ensures that evaluating the likelihood of $\vec w$ (and, hence, of $\vec Y$) will be computationally scalable. Because spatial covariances decrease with increasing distance, \citet{vecchia1988estimation} demonstrated that replacing the conditional set $\vec{w}_{1:(i-1)}$ by the smaller set of $m$ nearest neighbors (in terms of Euclidean distance) of $\vec{s}_i$ provides an excellent approximation to the conditional density in \eqref{eq:nngpin}. \citet{datta2016hierarchical} demonstrated that this is equivalent to $\vec A$ having at-most $m$ non-zero entries in each row and thereby corresponds to a proper probability distribution. Similarly, for prediction at a new location $\vec{s}$, a sparse $\vec{a}(\vec{s})$ in (\ref{eq:nngpout}) is constructed based on $m$-nearest neighbors of $\vec{s}$ among $\vec{s}_1, \dots, \vec{s}_N$. The resulting Gaussian Process is referred to as the Nearest Neighbor Gaussian Process (NNGP) and computation primarily involves small $m \times m$ matrix operations. Generalizing the use of nearest neighbors from expedient likelihood evaluations as in \citet{vecchia1988estimation} and \citet{stein2004approximating} to the well defined NNGP on the entire domain enables fully Bayesian inference and coherent recovery of the latent SREs. 

Using an NNGP prior for $Y(\vec{s}) - \vec X'(\vec{s})\vec \beta$, the model can be written as $\vec Y \sim N(\vec X\vec \beta, \vec {\widetilde \Sigma} (\vec \phi) )$ where $\vec {\widetilde \Sigma}$ is the NNGP covariance matrix derived from the full GP. A Bayesian specification is completed by specifying priors for the parameters $\vec \beta$ and $\vec \phi$. For this application, the covariance function $\mathbb C$ consists of an stationary exponential GP with variance $\sigma^2$ and range $\phi$ and a nugget process with variance $\sigma^2_\epsilon$ (see \eqref{YFRK}). We assign a normal prior for $\vec \beta$, inverse gamma priors for $\sigma^2_w$ and $\sigma^2_\epsilon$ and a uniform prior for $\phi$. A Gibbs sampler for the model involves conjugate updates for $\vec \beta$ and metropolis random walk updates for $\vec \phi = (\sigma^2_w, \sigma^2_\epsilon, \phi)'$. 

Letting $\alpha= \sigma^2_\epsilon/\sigma^2_w$, the model can also be expressed as $\vec Y \sim N(\vec X\vec \beta, \sigma^2_w \vec {\widetilde R } (\phi,\alpha) )$ where $\vec {\widetilde R }$ is the NNGP matrix derived from $\vec C(\phi) + \alpha \vec I$, $\vec C(\phi)$ being the correlation matrix of the exponential GP. Fixing $\alpha$ and $\phi$ gives a conjugate Normal-Inverse Gamma posterior distribution for $\vec \beta$ and $\sigma^2_w$. Predictive distributions for $y(s)$ at new locations can also be obtained as $t$-distributions. The fixed values of $\alpha$ and $\phi$ can be chosen from a grid-search by minimizing root mean square predictive error score based on $K$-fold cross validation. This hybrid approach departs from fully Bayesian philosophy by using hyper-parameter tuning. However, it offers a pragmatic solution for massive spatial datasets. We refer to this model as the {\em conjugate NNGP} model and the fully Bayesian approach described above as the {\em response NNGP} model. Detailed algorithms for both the models are provided in \citet{finley2017algorithm}. NNGP models for analyzing massive spatial data are available on CRAN as the R-package {\em spNNGP} \citep{spnngp}. 

\subsection{Stochastic PDEs}

The stochastic partial differential equation approach (SPDE) is based
on the equivalence between Mat\'ern covariance fields and stochastic
PDEs, in combination with the Markov property that on 2-dimensional
domains holds for integer valued smoothness parameters in the Mat\'ern
family. The starting point is a basis expansion for $w(\vec{s})$ of
the form \eqref{basis}, where the basis functions $h_k(\vec{s})$ are
chosen to be piecewise linear on a triangulation of the
domain~\citep{lindgren2011explicit}.  The optimal joint distribution
for the $\theta_k$ coefficients is obtained through a finite element
construction, which leads to a sparse inverse covariance matrix
(precision) $\vec{Q}_\theta(\vec{\phi})$. The precision matrix
elements are polynomials in the precision and inverse range parameters
($1/\phi_{\sigma}^2$ and $1/\phi_r$), with sparse matrix coefficients
that are determined solely by the choice of triangulation. This
differs from the sequential Markov construction of the NNGP method
which instead constructs a square-root free
$\vec{L}\vec{D}\vec{L}'$ Cholesky decomposition of its resulting
precision matrix (in a reverse order permutation of the elements).

The spatial process is specified through a joint Gaussian model for
$\vec{z}=(\vec{\theta},\, \vec{\beta})$ with prior mean $\vec{0}$ and
block-diagonal precision
$\vec{Q}_z=\diag(\vec{Q}_\theta,\vec{Q}_\beta)$, where 
$\vec{Q}_\beta=\vec{I}\cdot 10^{-8}$ gives a vague prior for
$\vec{\beta}$.  Introducing the sparse basis evaluation matrix
$\vec{H}$ with elements $H_{ij}=h_j(\vec{s}_i)$ and covariate matrix
$\vec{X}=X_j(\vec{s}_i)$, the observation model is then $\vec{Y} =
\vec{X}\vec{\beta} + \vec{H} \vec{\theta} + \vec{\epsilon}$.  The
design matrix for the joint vector $\vec{z}$ is
$\vec{A}=(\vec{H},\,\vec{X})$, and $\vec{\epsilon}$ is a zero mean
observation noise vector with diagonal precision
$\vec{Q}_\epsilon=\vec{I}/\sigma_\epsilon^2$.

Using the precision based equations for multivariate Normal
distributions, the conditional precision and expectation for $\vec{z}$
are given by $\vec{Q}_{z|y} = \vec{Q}_z + \vec{A'} \vec{Q}_\epsilon
\vec{A}$ and $\vec{\mu}_{z|y} = \vec{Q}_{z|y}^{-1} \vec{A'}
\vec{Q}_\epsilon \vec{Y}$, where sparse Cholesky factorisation of
$\vec{Q}_{z|y}$ is used for the linear solve. The elements of
$\vec{z}$ are automatically reordered to keep the Cholesky factors as
sparse as possible.  The resulting computational and storage cost for
the posterior predictions and multivariate Gaussian likelihood of a
spatial Gaussian Markov random field of this type with $K$ basis
functions is $\mathcal{O}(K^{3/2})$. Since the direct solver does not
take advantage of the stationarity of the model, the same prediction
cost would apply to non-stationary models.  For larger problems, more
easily parallelizeable iterative sparse solvers (e.g.\ multigrid) can
be applied, but for the relatively small size of the problem here, the
straightforward implementation of a direct solver is likely
preferable.  The posterior covariance elements of $\vec{Q}_{z|y}^{-1}$
corresponding to the non-zero structure of $\vec{Q}_{z|y}$ are
obtained through Takahashi recursions as a post-processing step on the
Cholesky factor of $\vec{Q}_{z|y}$~\citep[see][]{rue2009inla}. These
elements are precisely the ones needed to compute the final predictive
variances $\Var[\mu(\vec{s}_0) + w(\vec{s}_0) + \epsilon_0 \mid
  \vec{Y}]$ for each prediction location $\vec{s}_0$.

The triangulation nodes were here chosen to coincide with the
observation lattice, and in order to avoid unwanted boundary effects,
the triangulation extends a short distance outside the domain. This
extension has only a small effect on the computational cost, since the
triangles are allowed to be larger than inside the domain of interest,
and therefore the extension doesn't need as many nodes as in a regular
lattice extension.  In addition, the exponential covariance is a
Mat\'ern covariance with smoothness $0.5$, and hence is not Markovian
on $\mathbb{R}^2$.  Where the LK method approaches this by using a sum
of several Markovian components, the SPDE implementation in
\texttt{INLA}~\citep{RINLA} instead uses a parsimonious Markovian
spectral approximation for a single field. The resulting model is a
second order Markov random field on the coefficients
$\{\theta_k\}$. For details of the approximation see the authors'
response to the discussion of \citet{lindgren2011explicit}.

The implementation of the SPDE method used here is based on the
\texttt{R} package \texttt{INLA} \citep{RINLA}, which is aimed at
Bayesian inference for latent Gaussian models (in particular Bayesian
generalised linear, additive, and mixed models) using integrated
nested Laplace approximations \citep{rue2009inla}.  The parameter
optimization for
$\vec{\phi}=(\phi_{r},\phi_{\sigma},\sigma_\epsilon^2)$ uses general
numerical log-likelihood derivatives, thus the
full Bayesian inference was therefore turned off, leading to an
empirical Bayes estimate of the covariance parameters. Most of the
running time is still spent on parameter optimization, but using the
same parameter estimation technique as for LK, in combination with a
purely Gaussian implementation, substantively
reduces the total running time even without specialized code for the
derivatives.

\subsection{Periodic Embedding}
When the observation locations form a regular grid, and the model is stationary, methods that make use of the discrete Fourier transform (DFT), also known as spectral methods, can be statistically and computationally beneficial, since the DFT is an approximately decorrelating transform, and it can be computed quickly and with low memory burden using fast Fourier transform (FFT) algorithms. For spatially gridded data in two or higher dimensions--as opposed to time series data in one dimension--there are two prominent issues to be addressed. The first is edge effects, and the second is missing values. By projecting onto trigonometric bases, spectral methods essentially assume that the process is periodic on the observation domain, which leads to bias in the estimates of the spectrum \citep{guyon1982parameter,dahlhaus1987edge}. \cite{guinness2017circulant} and \cite{guinness2017spectral} propose the use of small domain expansions and imputing data in a periodic fashion on the expanded lattice. Imputation-based methods also solve the second issue of missing values, since the missing observations can be imputed as well.

The methods presented here follow the iterative semiparametric approach in \cite{guinness2017spectral}. \cite{guinness2017circulant} provides an alternative parametric approach. For this section, let $\vec{N} = (N_1,N_2)$ give the dimensions of the observation grid (in the surface temperature dataset $\vec{N} = (300,500)$). Let $\tau$ denote an expansion factor, and let $m = \lfloor \tau \vec{N} \rfloor$ denote the size of the expanded lattice. We use $\tau = 1.2$ in all examples, so that $m = (360,600)$ in the surface temperature dataset. Let $\vec{U}$ be the vector of observations, and $\vec{V}$ be the vector of missing values on the grid of size $m$, making the full vector $\vec{Y} = (\vec{U}',\vec{V}')'$. The discrete Fourier transform of the entire vector is
\begin{align*}
J(\vec{\omega}) = \frac{1}{\sqrt{m_1 m_2}} \sum_{\vec{s}} Y(\vec{s}) \exp(-i\vec{\omega}'\vec{s}),
\end{align*}
$\vec{\omega} = (\omega_1,\omega_2)'$ is a spatial frequency with $\omega_j \in [0,2\pi]$, $i = \sqrt{-1}$, and $\vec{\omega}'\vec{s} = \omega_1 s_1 + \omega_2 s_2$.

The procedure is iterative. At iteration $k$, the spectrum $f_k$ is updated with
\begin{align}\label{iterator}
f_{k+1}(\omega) = \sum_{\vec{\nu}} E_k( |J(\vec{\nu})|^2 \, | \, \vec{U} ) \alpha( \vec{\omega} - \vec{\nu} ),
\end{align}
where $\alpha$ is a smoothing kernel, and $E_k$ is expected value under the multivariate normal distribution with stationary covariance function
\begin{align*}
R_k( \vec{h} ) = \frac{1}{m_1 m_2} \sum_{\vec{\omega} \in \mathbb{F}_m } f_k(\vec{\omega}) \exp(i \vec{\omega} ' \vec{h}),
\end{align*}
where $\mathbb{F}_m$ is the set of Fourier frequencies on a grid of size $m$. This is critical since it ensures that $R_k$ is periodic on the expanded grid. In practice, the expected value in \eqref{iterator} is replaced with $|J(\vec{\nu})|^2$ computed using an imputed vector $\vec{V}$, a conditional simulation of missing values given $\vec{U}$ under covariance function $R_k$. This ensures that the imputed vector $\vec{V}$ is periodic on the expanded lattice and reduces edge effects. The iterative procedure can also be run with an intermediate parametric step in which the Whittle likelihood \citep{whittle1954stationary} is used to estimate a parametric spectral density, which is used to filter the imputed data prior to smoothing the spectrum. See \cite{guinness2017spectral} for details about more elaborate averaging schemes and monitoring for convergence of the iterative method.

\subsection{Metakriging}
Spatial metakriging is an approximate Bayesian method that is not tied to any specific model and is partly algorithmic in nature. In particular, any spatial model described above can be used to draw inference from subsets (as described below).  From \eqref{GaussPDF}, let the $N\times N$ covariance matrix be determined by a set of covariance parameters $\vec{\phi}$ such that $\bSigma = \bSigma(\vec{\phi})$ (e.g.\ $\vec{\phi}$ could represent decay parameters from the Mat\'{e}rn covariance function) and $\mu(\vec{s}) = \vec{X}'(\vec{s})\vec{\beta}$ where $\vec{X}(\vec{s})$ is a set of known covariates with unknown coefficients $\bbeta$.  Further, let the sampled locations $\mathcal{S}=\{\bs_1,...,\bs_N\}$ be partitioned into sets $\{\mathcal{S}_1,...,\mathcal{S}_K\}$ such that $\mathcal{S}_i\cap\mathcal{S}_j=\emptyset$ for $i\neq j$ and the corresponding partition of the data be given by $\{\by_k, \bX_k\}$, for $k=1,2,\ldots,K$, where each $\by_k$ is $n_k\times 1$ and $\bX_k$ is $n_k\times p$. 
Assume that we are able to obtain posterior samples for $\bOmega = \{\bbeta, \bphi\}$ from \eqref{GaussPDF} applied independently to each of $K$ subsets of the data in \textit{parallel on different cores}. To be specific, assume that $\bOmega_k = \{\bOmega_k^{(1)}, \bOmega_k^{(2)},\ldots, \bOmega_k^{(M)}\}$ is a collection of $M$ posterior samples from $p(\bOmega\given \by_k)$. We refer to each $p(\bOmega\given \by_k)$ as a ``subset posterior.'' The metakriging approach we outline below attempts to combine, optimally and meaningfully, these subset posteriors to arrive at a legitimate probability density. We refer to this as the ``metaposterior''.

Metakriging relies upon the unique geometric median (GM) of the subset posteriors \citep{minsker2014robust,minsker2015geometric}. We envision the individual posterior densities $p_k \equiv p(\bOmega\given \by_k)$ to be residing on a Banach space ${\cal H}$ equipped with norm $\|\cdot\|_{\rho}$. The GM is defined as
\begin{align}\label{Eq: GM}
\pi^*(\bOmega\given \by) = \arg\min\limits_{\pi\in\mathcal{H}}\sum_{k=1}^{K}\|p_k-\pi\|_{\rho}\;  ,
\end{align}
where $\by = (\by_1', \by_2',\ldots,\by_K')'$. The norm quantifies the distance between any two posterior densities $\pi_1(\cdot)$ and $\pi_2(\cdot)$ as $\|\pi_1 - \pi_2\|_{\rho} = \|\int \rho(\bOmega,\cdot)d(\pi_1-\pi_2)(\bOmega)\|$, where $\rho(\cdot)$ is a positive-definite kernel function. In what follows, we assume $\rho(z_1,z_2)=\exp(-||z_1-z_2||^2)$.

The GM is unique. Further, the geometric median lies in the convex hull of the individual posteriors, so $\pi^*(\bOmega\given \by)$ is a legitimate probability density. Specifically, $\pi^*(\bOmega\given \by)=\sum_{k=1}^{K}
\alpha_{\rho,k}(\by)p_k$, $\sum_{k=1}^{K}\alpha_{\rho,k}(\by)=1$, each $\alpha_{\rho,k}(\by)$ being a function of $\rho,\by$, so that $\int_{\bOmega}\pi^*(\bOmega\given \by)d\bOmega=1$.

Computation of the geometric median $\pi^*\equiv \pi^*(\bOmega\given \by)$ proceeds by employing the popular Weiszfeld's iterative algorithm that estimates $\alpha_{\rho,k}(\by)$ for every $k$ from the subset posteriors $p_k$. To further elucidate, we use a well known result that the geometric median $\pi^*$ satisfies, $$\pi^*=\frac{\sum_{k=1}^{K}\|p_k-\pi^*\|_{\rho}^{-1}p_k}{\sum_{k=1}^{K}\|p_k-\pi^*\|_{\rho}^{-1}}$$ so that $\alpha_{\rho,k}(\by)= \|p_k-\pi^*\|_{\rho}^{-1}/ \sum_{j=1}^{K}\|p_k-\pi^*\|_{\rho}^{-1}$.
Since there is no apparent closed form solution for $\alpha_{\rho,k}(\by)$ that satisfies this equation, one needs to resort to the Weiszfeld iterative algorithm outlined in \citet{minsker2014robust} to produce an empirical estimate of $\alpha_{\rho,k}(\by)$ for all $k=1,..,K$.

\cite{guhaniyogi2018meta} show that, for a large sample, $\pi^*(\cdot\given \by)$ provides desirable approximation of the full posterior distribution in certain restrictive settings. It is, therefore, natural to approximate the posterior predictive distribution $p(y(s_0)\given \by)$ by the subset posterior predictive distributions $p(y(s_0)\given \by_k)$. Let $\{y(s_0)^{(j,k)}\}_{j=1}^{M}$, $k=1,\ldots,K$, be samples obtained from the posterior predictive distribution $p(y(s_0)|\by_k)$ from the $k$-th subset posterior. Then,
\begin{align*}
p(y(s_0)\given \by)\approx \sum_{k=1}^K\alpha_{\rho,k}(\by)p(y(s_0)\given \by_k)=\sum_{k=1}^K\alpha_{\rho,k}(\by)\int p(y(s_0)\given \bOmega, \by_k)p(\bOmega\given \by_k)d\bOmega\; ,
\end{align*}
Therefore, the empirical posterior predictive distribution of the metaposterior is given by\\  $\sum_{k=1}^{K}\sum_{j=1}^{M}\frac{\alpha_{\rho,k}(\by)}{M}1_{y(s_0)^{(j,k)}}$, from which the posterior predictive median and the 95\% posterior predictive interval for the unobserved $y(s_0)$ are readily available. 

One important ingredient of spatial metakriging (SMK) is partitioning the dataset into subsets. For this article,
we adopt a random partitioning scheme that randomly divides data into $K=30$ exhaustive and mutually exclusive subsets.
The random partitioning scheme facilitates each subset to be a reasonable representative of the entire domain, so that each subset posterior acts as a ``weak learner" of the full posterior. We have explored more sophisticated partitioning schemes and found similar predictive inference.

For the sake of definiteness, this article uses the Gaussian process model for each subset inference which may lead to higher run time. However, the metakriging approach lends much more scalability when any of the above models is employed in each subset. In fact, ongoing research in spatial metakriging includes distributed spatial kriging (DISK) \citep{guhaniyogi2017divide} which scales the 
modified predictive process to millions of observations.


\subsection{Gapfill}
The gapfill method \citep{Gerb:etal:16} differs from the other herein presented methods in that it is purely algorithmic, distribution-free, and, in particular, not based on Gaussian processes.  Like other prediction methods popular within the satellite imaging community (see \citealt{Gerb:etal:16} and \citealt{Weis:etal:14} for reviews), the gapfill method is attractive because of its low computational workload.  A key aspect of gapfill is that it is designed for parallel processing, which allows the user to exploit computing resources at different scales including large servers.  Parallelization is enabled by predicting each missing value separately based on only a subset of the data.

To predict the value $Y(\vec{s}_0)$ at location $\vec{s}_0$ gapfill first selects a suitable subset $\vec{A}=\{Y(\vec{s}_i): \vec{s}_i \in \N(\vec{s}_0)\}$, where $\N(\vec{s}_0)$ defines a spatial neighborhood around $\vec{s}_0$.  Finding $\vec{A}$ is formalized with rules, which reassure that $\vec{A}$ is small but contains enough observed values to inform the prediction. In this study, we require $\vec{A}$ to have an extent of at least $5\times 5$ pixels and to contain at least $25$ non-missing values.  Subsequently, the prediction of $Y(\vec{s}_0)$ is based on $\vec{A}$ and relies on sorting algorithms and quantile regression. Moreover, prediction intervals are constructed using permutation arguments (see \citealt{Gerb:etal:16} for more details on the prediction and uncertainty intervals).

The gapfill method was originally designed for spatio-temporal data, in which case the neighborhood $\N(\vec{s}_0)$ is defined in terms of the spatial and temporal dimensions of the data. As a consequence, the implementation of gapfill in the R package \texttt{gapfill} \citep{gapfill} requires multiple images to work properly.  To mimic this situation, we shift the given images by one, two, and three pixels in both directions along the $x$ and $y$-axes.  Then the algorithm is applied to those $13$ images in total (one original image and $12$ images obtained through shifts of the original image).

\subsection{Local Approximate Gaussian Processes}
The local approximate Gaussian process \citep[{\tt laGP},][]{gramacy2015local}
addresses the big-$N$ problem in GP regression by taking a so-called {\em
transductive} approach to learning, where the fitting scheme is tailored to
the prediction problem \citep{vapnik:1995} as opposed to the usual
{\em inductive} approach of fitting first and predicting later conditional on
the fit.  A special case of {\tt laGP}, based on nearest neighbors, is simple
to describe. In order to predict at $\vec{s}$, simply train a Gaussian process predictor on the
nearest $m$ neighbors to $\vec{s}$; i.e.,\ use the data subset $\mathcal{Y}_m = \{Y(\vec{s}_i): \vec{s}_i \in \mathcal{N}_m(\vec{s})\}$,
where $\mathcal{N}_m(\vec{s})$ are the $m$ closest observed locations to $\vec{s}$ in terms
of Euclidean distance. If the data-generating mechanism is not at odds with
modeling assumptions (e.g., having a well-specified covariance structure),
then one can choose $m$ to be as large as possible, up to computational
limitations, in order to obtain an accurate approximation. Observe that this
use of nearest neighbors (NNs) for prediction is more akin to the classical
statistical/machine learning variety, in contrast to their use in determining
the global (inverse) covariance structure as described in Section \ref{NNGP}.

Interestingly, NNs do not comprise an optimal data subset for prediction
under the usual criteria such as mean-squared error. However, finding the
best $m$ of $N!/(m!(N-m)!)$ possible choices represents a combinatorially huge search. The {\tt
laGP} method generalizes this so-called nearest neighbor prediction algorithm (whose modern form
in spatial statistical literature is described by \citealt{emery:2009}) by
approximating that search with a greedy heuristic. First, start with a NN set
$\mathcal{Y}_{m_0}(\vec{s})= \{Y(\vec{s}_i): \vec{s}_i \in \mathcal{N}_{m_0}(\vec{s}))$ where $m_0 < m$, and then for $j=m_0+1,\dots,m$ successively choose $\vec{s}_{j}$ to
augment $\mathcal{Y}_{m_0}$ building up a local design data set one point at a time according to one of several
simple objective criteria related to mean-square prediction error. The idea is
to repeat in this way until there are $m$ observations in $\mathcal{Y}_m(\vec{s})$. \citeauthor{gramacy2015local}'s preferred
variation targets $\vec{s}_{j}$ which maximizes the {\em reduction} in predictive
variance at $\vec{s}$. To recognize a similar {\em global} design criterion called
{\em active learning Cohn} \citep{cohn:1996}, they dubbed this criterion ALC.
Qualitatively, these local ALC designs tend to have a cluster of neighbors and
``satellite'' points and have been shown to offer demonstrably better predictive
properties than NN and even full-data alternatives especially when the data
generating mechanism is at odds with the modeling assumptions. The reason is
that local fitting offers a way to cope with a certain degree of non-stationarity which
is common in many real data settings.

ALC search iterations and GP updating considerations as designs are built up,
are carefully engineered to lead to a method whose computations are of
$\mathcal{O}(N^3)$ complexity (i.e.,\ the same as the simpler NN alternative).
A relatively modest local design size of $m=50$ typically works well.
Moreover, calculations for each $\vec{s}$ are statistically independent of the
next, which means that they can be trivially parallelized. Through a cascade of
multi-core, multi-node and GPU parallelization, \citet{gramacy:niemi:weiss:2014}
and \citet{gramacy2015speeding} illustrated how $N$ in the millions, in terms
of both training and testing data sizes could be handled (and yield accurate
predictors) with less than an hour of computing time.  The {\tt laGP} method
has been packaged for {\sf R} and is available on CRAN \citep{gramacy2016laGP}.
Symmetric multi-core parallelization (via {\tt OpenMP}) and multi-node
automations (via the built-in {\tt parallel} package) work out-of-the box. GPU
extensions are provided in the source code but require custom compilation.  

A disadvantage to local modeling in this fashion is that a global predictive
covariance is unavailable.  Indeed, the statistically independent nature of
calculation is what makes the procedure computationally efficient and
parallelizable.  In fact, the resulting global predictive surface, over a
continuum of predictive $\vec{s}$-locations, need not even be smooth. However in
most visual representations of predictive surfaces it can be difficult to
distinguish between a genuinely smooth surface and what is plotted via the
{\tt laGP} predictive equations (see Figures \ref{simresfig} and \ref{satresfig} below).  Finally, it is worth noting that although
{\tt laGP} is applied here in a spatial modeling setting (i.e.,\ with two input
variables), it was designed for computer simulation modeling and has been shown
to work well in input dimension as high as ten.

\section{The Competition}\label{competition}
At the initial planning phase of this competition, we desired to compare a broad variety of approaches: from frequentist to Bayesian and from well-established to modern developments.  In accordance with this plan, efforts were made to contact a variety of research groups with strong expertise in a method to analyze the datasets.  After this outreach period, the research teams listed in Table \ref{groups} agreed to participate and implement their associated method.
\begin{table}[tb]
\caption{Research groups participating in the competition along with their selected method (competitor).}
\label{groups}
\begin{center}
\begin{tabular*}{\textwidth}
	{@{\extracolsep{\fill}} ll}
\hline \hline
Group Members & Method \\ \hline
Abhirup Datta \& Andrew Finley & Nearest Neighbor Processes \\
 Andrew~Finley & Predictive Processes \\
  Reinhard Furrer & Covariance Tapering \\
Florian Gerber & Gapfill \\
  Raj~Guhaniyogi & Metakriging \\
Matthew~J.~Heaton  & Spatial Partitioning \\
Andrew Zammit-Mangion & Fixed rank kriging \\
Matthias~Katzfuss \& Dorit~Hammerling & Multiresolution Approximations \\
 Finn~Lindgren & Stochastic Partial Differential Equations \\
 Joseph Guinness & Periodic Embedding \\
Douglas~Nychka & Lattice Kriging \\
Robert Gramacy \& Furong Sun & Local Approximate Gaussian Processes \\
\hline \hline
\end{tabular*}
\end{center}
\end{table}

Each group listed in Table \ref{groups} were provided with two training datasets: one real and one simulated.  The simulated dataset then represented a case where the covariance function was specified correctly while the real dataset represented a scenario where the covariance function was misspecified.  Both datasets consisted of observations on the same 500$\times$300 grid ranging longitude values of $-95.91153$ to $-91.28381$ and latitude values of $34.29519$ to $ 37.06811$.  The real dataset consisted of daytime land surface temperatures as measured by the Terra instrument onboard the MODIS satellite on August 4, 2016 (Level-3 data).  The data was downloaded from the MODIS reprojection tool web interface (MRTweb) located at \url{https://mrtweb.cr.usgs.gov/} and is provided as supplementary material to this article.  The latitude and longitude range, as well as the date, were chosen because of the sparse cloud cover over the region on this date (rather than by scientific interest in the date itself).  Namely, only 1.1\% of the Level-3 MODIS data were corrupted by cloud cover leaving 148,309/150,000 observed values to use for our purposes.
\begin{figure}
\centering
\includegraphics[scale=.5,angle=270]{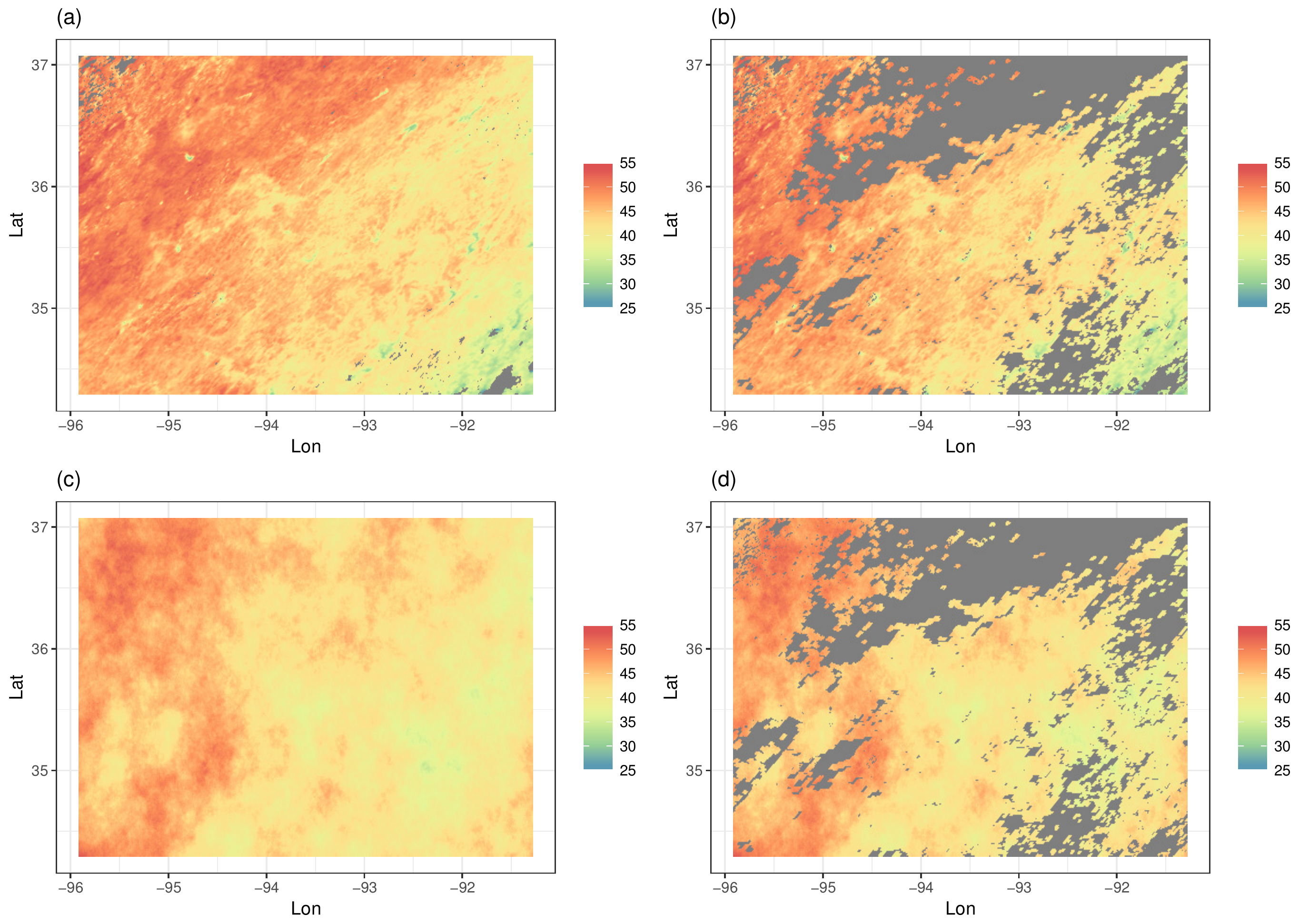}
\caption{The top row displays the (a) full and (b) training satellite datasets.  The bottom row displays the (c) full and (d) training simulated data.}
\label{datafig}
\end{figure}

The simulated dataset was created by, first, fitting a Gaussian process model with constant mean, exponential covariance function and a nugget effect to a random sample of 2500 observations from the above MODIS data. The resulting parameter estimates were then used to simulate 150,000 observations on the same grid as the MODIS data.  

In order to ensure a realistic analysis scenario, the missing data pattern on August 6, 2016 from the same MODIS satellite data product was used to separate each dataset into training and test sets.  After the split, the training set for the MODIS data consisted of 105,569 observations leaving 42,740 observations in the test set.  The training set for the simulated data also consisted of 105,569 observations but a test set size of 44,431 (the difference in test set size is contributed to missing data due to cloud cover in the original MODIS data).  Research teams were provided with the training set and the locations of the test set (but not the actual observation in the test set).  Figure \ref{datafig} displays the full datasets along with the corresponding training set provided to each research group.  All datasets used in this article are provided as supplementary material to this article.

Each group independently wrote code (all of which is included as supplementary material to this article) that provided (i) a point prediction for each location in the test set, (ii) a 95\% prediction interval for location in the test set or a corresponding standard error for the prediction, (iii) the average time required to implement the method per iteration and (iv) the total clock time needed to implement the method.  In order to minimize the number of confounding factors in this competition, each group was instructed to use an exponential correlation function (if applicable to their chosen method) and a nugget variance.  For the simulated data the groups were instructed to only use a constant mean (because this was how the data was originally simulated).  However, for the satellite data, the groups used a linear effect for latitude and longitude so that the residual process more closely resembled the exponential correlation.  The code from each team was then run on the Becker computing environment (256 GB of RAM and 2 Intel Xeon E5-2680 v4 \@ 2.40GHz CPUs with 14 cores each and 2 threads per core - totaling 56 possible threads for use in parallel computing) located at Brigham Young University (BYU).  Each team's code was run individually and no other processes were simultaneously run so as to provide an accurate measure of computing time.

Each method was compared in terms of mean absolute error ($\text{MAE} = n_{\text{test}}^{-1}\sum_{i=1}^{n_\text{test}} |y(\vec{s}_i) - \hat{y}(\vec{s}_i)|$), root mean squared error ($\text{RMSE} = (n_{\text{test}}^{-1}\sum_{i=1}^{n_\text{test}} (y(\vec{s}_i) - \hat{y}(\vec{s}_i))^2)^{1/2}$), continuous rank probability score (CRPS; see \citealt{gneiting2007strictly,gneiting2014probabilistic}), interval score (INT; see \citealt{gneiting2007strictly}) and prediction interval coverage (CVG; the percent of intervals containing the true value).  To calculate the CRPS, we assumed the associated predictive distribution was well approximated by a Gaussian distribution with mean centered at the predicted value and standard deviation equal to the predictive standard error.  In cases where only a prediction interval was provided, the predictive standard error was taken as $(U-L)/(2\times\Phi^{-1}(0.975))$ where $U$ and $L$ are the upper and lower ends of the interval, respectively.

\section{Competition Results}\label{results}

\subsection{Results for Simulated Data}
\begin{figure}
\centering
\includegraphics[scale=0.9,angle=270]{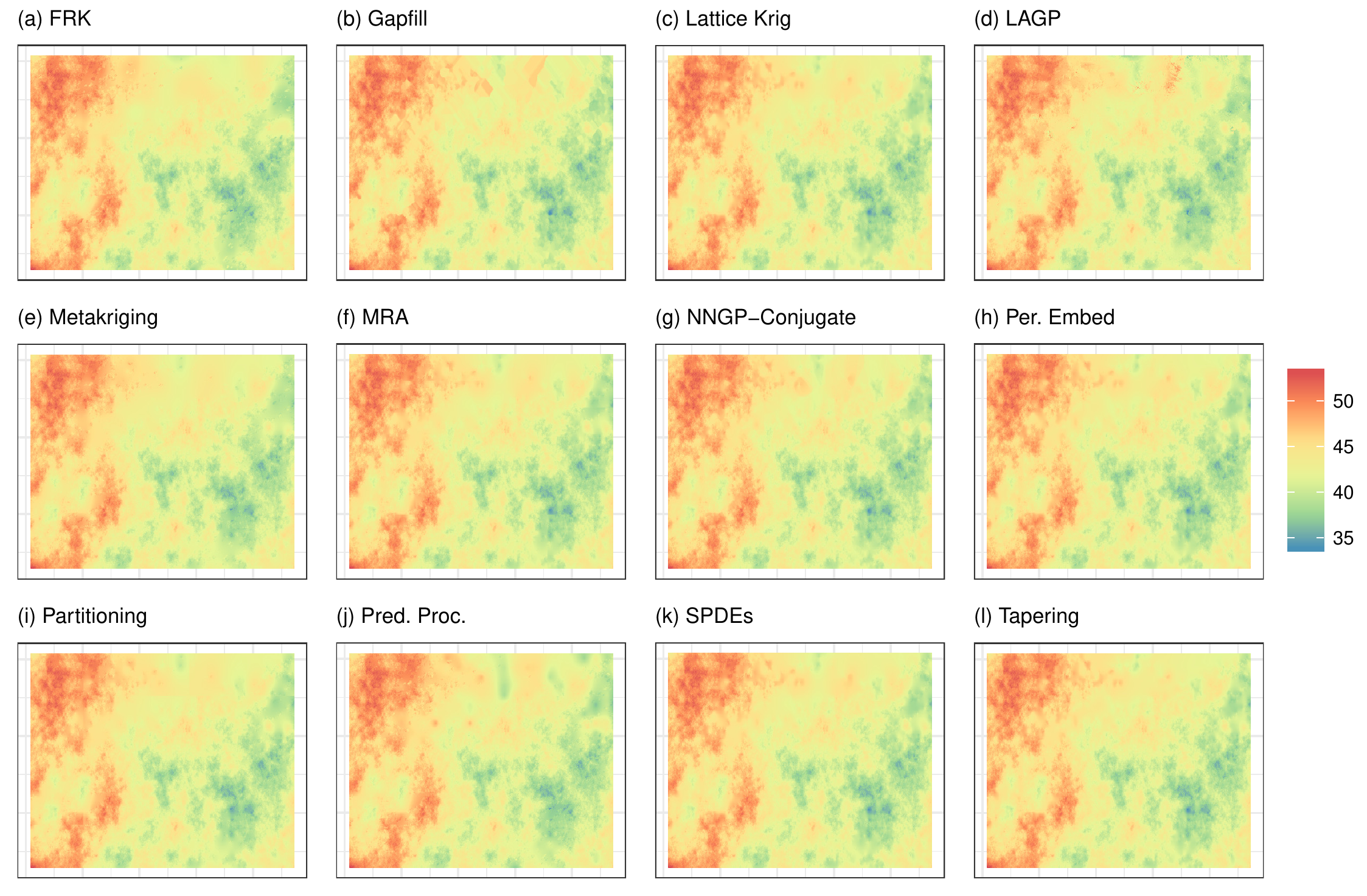}
\caption{Predictions for the simulated data using each of the competing methods.}
\label{simresfig}
\end{figure}

The numerical results for the simulated data competition are displayed in Table \ref{simnumres} and the associated predicted surfaces for each method are shown in Figure \ref{simresfig}.  First, consider the predictive accuracy as measured by the MAE and RMSE in Table \ref{simnumres}.  In terms of predictive accuracy, each method performed extremely well with the best MAE being 0.61 while the worst was only 1.03.  Similarly, the best RMSE was 0.83 compared to a worst RMSE of only 1.31.  Considering the range of the simulated data was $53.80-33.91=19.89$, a RMSE of 1.31 is quite accurate.

\begin{table}[tb]
\centering
\caption{Numerical scoring for each competing method on the simulated data. The best result of each score is bolded.}
\label{simnumres}
\begin{tabular*}{\textwidth}
	{@{\extracolsep{\fill}} lrrrrrrr}
  \hline \hline
 Method & MAE & RMSE & CRPS & INT & CVG & Run Time (Min) & Cores Used \\ 
  \hline
FRK & 1.03 & 1.31 & 0.74 & 8.35 & 0.84 & 2.18 & 1 \\ 
  Gapfill & 0.73 & 1.00 & 0.64 & 18.01 & 0.44 & 0.63 & 40 \\ 
  Lattice Krig & 0.63 & 0.87 & 0.45 & 4.04 & 0.97 & 25.58 & 1 \\ 
  LAGP &  0.79 & 1.11 & 0.57 & 5.71 & 0.90 & 2.28 & 40\\
  Metakriging & 0.74 & 0.97 & 0.53 & 4.69 & 0.99 & 2888.89 & 30 \\ 
  MRA & \textbf{0.61} & \textbf{0.83} & \textbf{0.43} & \textbf{3.64} & 0.93 & 13.57 & 1 \\ 
  NNGP Conjugate & 0.65 & 0.88 & 0.46 & 3.79 & \textbf{0.96} & 1.99 & 10 \\
  NNGP Response & 0.65 & 0.88 & 0.46 & 3.81 & \textbf{0.96} & 45.06 & 10 \\ 
  Partition & 0.64 & 0.86 & 0.47 & 5.05 & 0.86 & 77.56 & 55 \\ 
  Pred. Proc. & 0.89 & 1.21 & 0.79 & 12.75 & 0.77 & 639.23 & 1 \\ 
  SPDE & 0.62 & 0.86 & 0.59 & 7.81 & 1.00 & 138.34 & 2 \\ 
  Tapering & 0.69 & 0.97 & 0.55 & 6.39 & 1.00 & 188.36 & 1 \\ 
  Periodic Embedding & 0.65 & 0.91 & 0.47 & 4.16 & 0.97 & 13.31 & 1 \\
   \hline \hline
\end{tabular*}
\end{table}

While all the methods performed well in terms of predictive accuracy, when considering uncertainty quantification (UQ) some of the methods fared better than others.  For example, LatticeKrig, LAGP, metakriging, MRA, periodic embedding and NNGP all achieved near the nominal 95\% coverage rate.  In contrast, FRK, Gapfill, partitioning and PP achieved lower than nominal coverage while SPDE and tapering have higher than nominal coverage.  Considering UQ further, Gapfill and PP have large interval scores suggesting possible wide predictive intervals in addition to the penalty incurred from missing the true value.  In this regard, it is important to keep in mind that LAGP, metakriging, MRA, NNGP and PP all can specify the ``correct'' exponential correlation function.  Additionally, LK and SPDE have settings that can approximate the exponential correlation function well.  In contrast, some methods such as FRK and Gapfill are less suited to model fields with exponential correlation functions, which may partially explain their relatively poor prediction or coverage performance in this instance.


Finally, Figure \ref{simresfig} displays the predictive surfaces for each method on the simulated data.  The visual inspection of the predictive surfaces provides interesting insights into the various features of each method.  For example, because the Gapfill method was primarily designed for spatio-temporal data, we shifted the images to create ``pseudo'' datasets for the Gapfill algorithm.  However, this shifting resulted in a ``smeared'' pattern in the predictive surface which we hypothesize would not occur in the space-time setting.  Likewise, arguments by \citet{simpson2012order} and \citet{stein2014limitations} suggest that low rank methods oversmooth the data and such possible oversmoothing is seen in the predictive surfaces for FRK and PP.

\subsection{Results for Real Data}
The results for the real MODIS data are displayed in Table \ref{satnumres} and largely reiterate the results from the simulated data.  Namely, each method performed very well in terms of predictive accuracy.  The largest RMSE was only 2.52 which, when considered on the data range of $55.41-24.37=31.04$, is very small.  We note that, under the setup of the competition, some of the methods were forced to approximate a GP with isotropic exponential covariance function, which is the true covariance function of the simulated data, but most certainly not for the real data. Thus, the scores are lowest for those approximations that happened to result in a good fit to the data and not necessarily lowest for those methods that best approximated the exponential covariance.
\begin{table}[tb]
\centering
\caption{Numerical scoring for each competing method on the satellite data. The best result of each score is bolded.}
\label{satnumres}
\begin{tabular*}{\textwidth}
	{@{\extracolsep{\fill}} lrrrrrrr}
  \hline \hline
 Method & MAE & RMSE & CRPS & INT & CVG & Run Time (Min) & Cores Used \\ 
  \hline
FRK & 1.96 & 2.44 & 1.44 & 14.08 & 0.79 & 2.32 & 1\\ 
   Gapfill & 1.33 & 1.86 & 1.17 & 34.78 & 0.36 & 1.39 & 40 \\ 
   Lattice Krig & 1.22 & 1.68 & 0.87 & 7.55 & 0.96 & 27.92 & 1 \\ 
   LAGP & 1.65 & 2.08 & 1.17 & 10.81 & 0.83 & 2.27 & 40 \\
   Metakriging & 2.08 & 2.50 & 1.44 & 10.77 & 0.89 & 2888.52 & 30 \\ 
   MRA & 1.33 & 1.85 & 0.94 & 8.00 & 0.92 & 15.61 & 1\\ 
   NNGP Conjugate & 1.21 & 1.64 & 0.85 & 7.57 & \bf{0.95} & 2.06 & 10 \\
   NNGP Response & 1.24 & 1.68 & 0.87 & 7.50 & 0.94 & 42.85 & 10 \\ 
   Partition & 1.41 & 1.80 & 1.02 & 10.49 & 0.86 & 79.98 & 55 \\ 
  Pred. Proc. & 2.05 & 2.52 & 1.85 & 26.24 & 0.75 & 640.48 & 1 \\ 
  SPDE & \bf{1.10} & \bf{1.53} & \bf{0.83} & 8.85 & 0.97 & 120.33 & 2 \\ 
   Tapering & 1.87 & 2.45 & 1.32 & 10.31 & 0.93 & 133.26 & 1 \\ 
   Periodic Embedding & 1.29 & 1.79 & 0.91 & \bf{7.44} & 0.93 & 9.81 & 1 \\
   \hline \hline
\end{tabular*}
\end{table}

The largest discrepancies among the competing methods is again in terms of uncertainty quantification.  Lattice kriging, metakriging, MRA, NNGP and periodic embedding again achieved near nominal coverage rates with small interval scores and CRPS.  The SPDE and tapering approaches did better in terms of coverage in that the empirical rates were near nominal (recall that the corresponding coverage rates were too high for the simulated data for these methods).  In contrast, the coverage rates on the MODIS data for FRK, Gapfill, LAGP, partitioning and predictive processes were too small resulting in larger interval scores.

\begin{figure}
\centering
\includegraphics[scale=.9,angle=270]{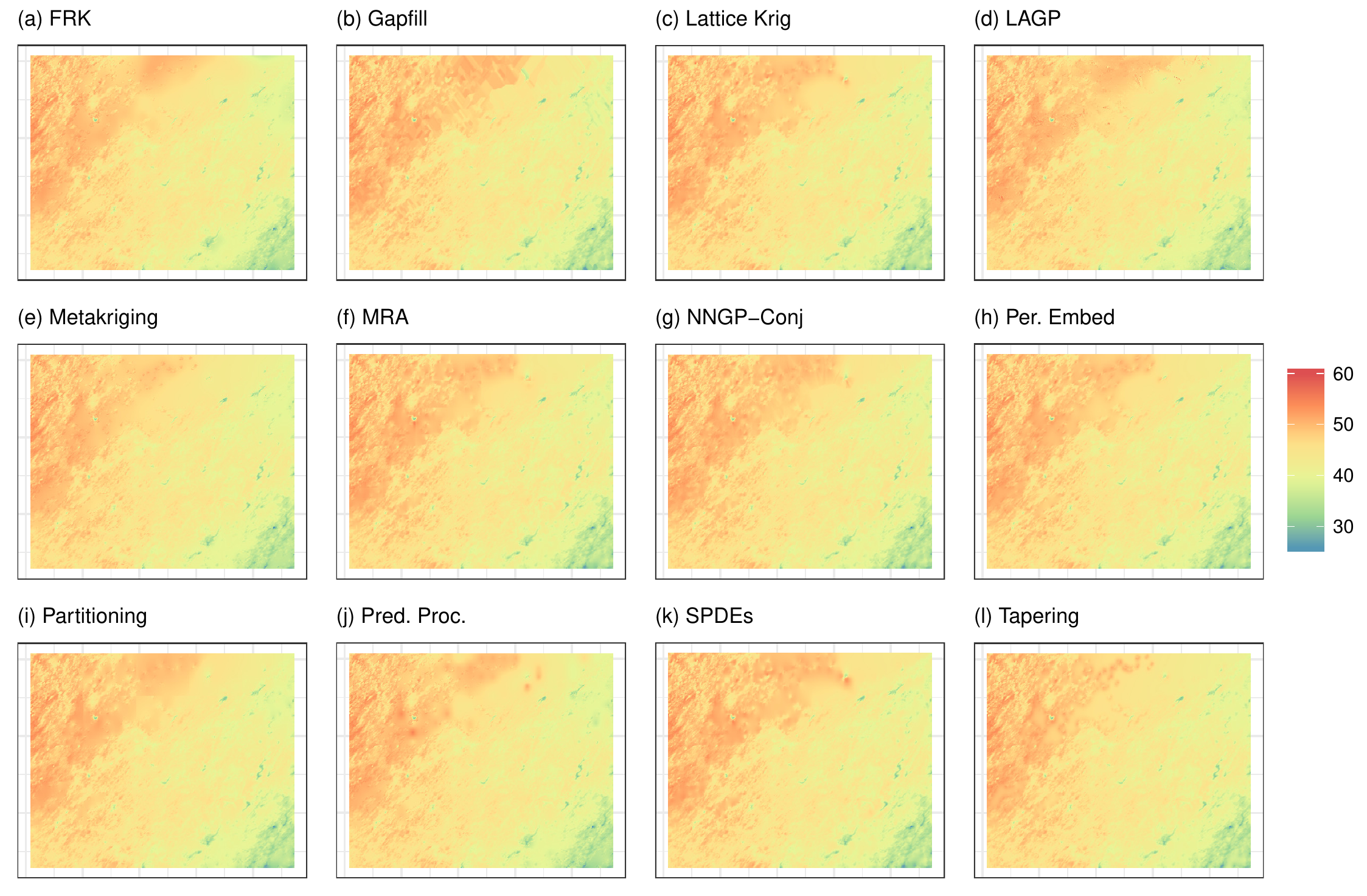}
\caption{Predictions for the satellite data using each of the competing methods.}
\label{satresfig}
\end{figure}

Finally, visual inspections of the predictive surfaces for the MODIS data are shown in Figure \ref{satresfig}.  Notably the majority of the methods smooth out the predictions in the north-central region.  This is to be expected because such predictions are considered ``long-range'' with very little (or no) observed data in this region (see Figure \ref{datafig}).  Hence, predictions for this region rely more heavily on the overall mean surface rather than borrowing information from neighboring observations (of which there is none).  Again, the ``shifting'' used for the Gapfill algorithm is again apparent in the predictive surface.  As with the simulated data, we hypothesize that such ``smeared'' predictive surfaces for Gapfill would not occur under the spatio-temporal setting.

\section{Conclusions}\label{conc}
The contribution of this article was four-fold: (i) provide an overview of the plethora of methods available for analyzing large spatial datasets, (ii) provide a brief comparison of the methods by implementing a case study competition among research groups, (iii) make available the code to analyze the data to the broader scientific community and (iv) provide an example of the common task framework for future studies to follow when comparing various analytical methods.  In terms of comparison, each of the methods performed very well in terms in predictive accuracy suggesting that any of the above methods are well suited to the task of prediction.  However, the methods differed in terms of their ability to accurately quantify the uncertainty associated with the predictions.  While we saw that some methods did consistently well in both predictive performance and nominal coverage on the simulated and real data, in general we can expect performance of any method to change with size of the dataset, measurement error variance, and the nature of missingness.  However, the data scenario's considered here are relatively representative of a typical spatial analysis such that our results can be used as a guide for practitioners. 

At the outset of this study, run time and computation time for each method was of interest.  However, because many of these methods are very young in their use and implementation, the variability across run time was too great to be used as a measure to compare the methods.  For example, some methods are implemented in R while others are implemented in MATLAB.  Still, others use R as a front end to call C-optimized functions.  Hence, while we reported the run times in the results section, we provide these as more of an ``off the shelf'' run time estimate rather than an optimized run time.  Until time allows for each method to be further developed and software becomes available comparing run times can be misleading.

Importantly, no effort was made to standardize the time spent on this project by each group.  Some groups were able to quickly code up their analysis from existing R or MATLAB libraries.  Others, however, had to spend more time writing code specific to this analysis.  Undoubtedly, some groups likely spent more time running ``in house'' cross-validation studies to validate their model predictions prior to the final run on the BYU servers while others did not.  Because of this difference, we note that some of the discrepancies in results seen here may be attributable to the amount of effort expended by each group.  However, we still feel that the results displayed herein give valuable insight into the strengths and weaknesses of each method.

This study, while thorough, is non-comprehensive in that other methods for large spatial data \citep[e.g.][]{sang2012full,stein2013stochastic,kleiber2015equivalent, castrillon2016multi,sun2016statistically,litvinenko2017likelihood} were not included.  Additionally, methods are sure to be developed in the future which are also viable for modeling large spatial data \citep[see][]{ton2017spatial,taylor2018spatial}.  We made attempts to invite as many groups as possible to participate in this case study but, due to time and other constraining factors, not all groups were able to participate.  However, in our opinion, the methods compared herein are representative of the most common methods for large spatial data at the time of writing.

We note that the data scenarios considered in this case study do not cover the spectrum of issues related to spatial data.  That is, spatial data may exhibit anisotropy, nonstationarity, large and small range spatial dependence as well as various signal-to-noise ratios.  Hence, we note that further practical distinctions between these various methods could be made depending on their applicability to these various spatial data scenarios.  However, the comparison included here serves as a nice baseline case for method performance.  Further research can develop case study competitions for these more complicated scenarios.

Notably, each method was compared only in terms of predictive accuracy.  Further comparisons could include estimation of underlying model parameters.  The difficulty in comparing estimation, however, is that not all the methods use the same model structure.  For example, NNGP uses an exponential covariance while Gapfill does not require a specified covariance structure.  Hence, we leave the comparison of the parameter estimates to a future study.

This comparison focused solely on spatial data.  Hence, we stress that the results found here are applicable only to the spatial setting.  However, spatio-temporal data are often considerably larger and more complex than spatial data.  Many of the above methods have extensions to the space time setting (e.g.,\ Gapfill is built directly for spatio-temporal settings).  Further research is needed to compare these methods in the spatio-temporal setting.  

\section*{Acknowledgements}
This material is based upon work supported by the National Science Foundation (NSF) under Grant Number DMS-1417856.  Dr.\ Katzfuss is partially supported by NSF Grants DMS--1521676 and DMS--1654083.  Dr.\ Gramacy and Furong Sun are partially supported by NSF Award \#1621746.  Dr.\ Finley was partially supported by NSF DMS-1513481, EF-1241874, EF-1253225, and National Aeronautics and Space Administration (NASA) Carbon Monitoring System (CMS) grants. Any opinions, findings, and conclusions or recommendations expressed in this material are those of the author(s) and do not necessarily reflect the views of the NSF or NASA.

\begin{supplement}[id=suppA]
\sname{Supplement A}
\stitle{Computational Details}
\slink[doi]{TBD}
\sdatatype{.pdf}
\sdescription{Additional details regarding the implementation of some of the methods to the training datasets.}
\end{supplement}

\begin{supplement}[id=suppB]
\sname{Supplement B}
\stitle{Code and Data}
\slink[doi]{TBD}
\sdatatype{.zip}
\sdescription{The code used to implement each method along with the datasets.  Code to reproduce the results in the paper (e.g. Tables 2 and 3) is also provided.}
\end{supplement}

\bibliographystyle{./imsart-nameyear}
\bibliography{References}

\begin{thebibliography}{85}

\bibitem[\protect\citeauthoryear{Anderson, Lee and
  Dean}{2014}]{anderson2014identifying}
\begin{barticle}[author]
\bauthor{\bsnm{Anderson},~\bfnm{Craig}\binits{C.}},
  \bauthor{\bsnm{Lee},~\bfnm{Duncan}\binits{D.}} \AND
  \bauthor{\bsnm{Dean},~\bfnm{Nema}\binits{N.}}
(\byear{2014}).
\btitle{Identifying clusters in Bayesian disease mapping}.
\bjournal{Biostatistics}
\bvolume{15}
\bpages{457--469}.
\end{barticle}
\endbibitem

\bibitem[\protect\citeauthoryear{Banerjee, Carlin and
  Gelfand}{2014}]{banerjee2014hierarchical}
\begin{bbook}[author]
\bauthor{\bsnm{Banerjee},~\bfnm{Sudipto}\binits{S.}},
  \bauthor{\bsnm{Carlin},~\bfnm{Bradley~P}\binits{B.~P.}} \AND
  \bauthor{\bsnm{Gelfand},~\bfnm{Alan~E}\binits{A.~E.}}
(\byear{2014}).
\btitle{Hierarchical modeling and analysis for spatial data}.
\bpublisher{Crc Press}.
\end{bbook}
\endbibitem

\bibitem[\protect\citeauthoryear{Banerjee et~al.}{2008}]{banerjee2008gaussian}
\begin{barticle}[author]
\bauthor{\bsnm{Banerjee},~\bfnm{Sudipto}\binits{S.}},
  \bauthor{\bsnm{Gelfand},~\bfnm{Alan~E}\binits{A.~E.}},
  \bauthor{\bsnm{Finley},~\bfnm{Andrew~O}\binits{A.~O.}} \AND
  \bauthor{\bsnm{Sang},~\bfnm{Huiyan}\binits{H.}}
(\byear{2008}).
\btitle{Gaussian predictive process models for large spatial data sets}.
\bjournal{Journal of the Royal Statistical Society: Series B (Statistical
  Methodology)}
\bvolume{70}
\bpages{825--848}.
\end{barticle}
\endbibitem

\bibitem[\protect\citeauthoryear{Barbian and
  Assun{\c{c}}{\~a}o}{2017}]{barbian2017spatial}
\begin{barticle}[author]
\bauthor{\bsnm{Barbian},~\bfnm{M{\'a}rcia~H}\binits{M.~H.}} \AND
  \bauthor{\bsnm{Assun{\c{c}}{\~a}o},~\bfnm{Renato~M}\binits{R.~M.}}
(\byear{2017}).
\btitle{Spatial subsemble estimator for large geostatistical data}.
\bjournal{Spatial Statistics}
\bvolume{22}
\bpages{68--88}.
\end{barticle}
\endbibitem

\bibitem[\protect\citeauthoryear{Bevilacqua et~al.}{2016}]{Bevi:etal:16}
\begin{bmisc}[author]
\bauthor{\bsnm{Bevilacqua},~\bfnm{M.}\binits{M.}},
  \bauthor{\bsnm{Faouzi},~\bfnm{T.}\binits{T.}},
  \bauthor{\bsnm{Furrer},~\bfnm{R.}\binits{R.}} \AND
  \bauthor{\bsnm{Porcu},~\bfnm{E.}\binits{E.}}
(\byear{2016}).
\btitle{Estimation and Prediction using Generalized {W}endland Covariance
  Function under Fixed Domain Asymptotics}.
\bnote{arXiv:1607.06921v2}.
\end{bmisc}
\endbibitem

\bibitem[\protect\citeauthoryear{Bradley et~al.}{2016}]{bradley2016comparison}
\begin{barticle}[author]
\bauthor{\bsnm{Bradley},~\bfnm{Jonathan~R}\binits{J.~R.}},
  \bauthor{\bsnm{Cressie},~\bfnm{Noel}\binits{N.}},
  \bauthor{\bsnm{Shi},~\bfnm{Tao}\binits{T.}} \betal{et~al.}
(\byear{2016}).
\btitle{A comparison of spatial predictors when datasets could be very large}.
\bjournal{Statistics Surveys}
\bvolume{10}
\bpages{100--131}.
\end{barticle}
\endbibitem

\bibitem[\protect\citeauthoryear{Castrillon-Cand{\'a}s, Genton and
  Yokota}{2016}]{castrillon2016multi}
\begin{barticle}[author]
\bauthor{\bsnm{Castrillon-Cand{\'a}s},~\bfnm{Julio~E}\binits{J.~E.}},
  \bauthor{\bsnm{Genton},~\bfnm{Marc~G}\binits{M.~G.}} \AND
  \bauthor{\bsnm{Yokota},~\bfnm{Rio}\binits{R.}}
(\byear{2016}).
\btitle{Multi-level restricted maximum likelihood covariance estimation and
  kriging for large non-gridded spatial datasets}.
\bjournal{Spatial Statistics}
\bvolume{18}
\bpages{105--124}.
\end{barticle}
\endbibitem

\bibitem[\protect\citeauthoryear{Cohn}{1996}]{cohn:1996}
\begin{binproceedings}[author]
\bauthor{\bsnm{Cohn},~\bfnm{D.~A.}\binits{D.~A.}}
(\byear{1996}).
\btitle{Neural Network Exploration Using Optimal Experimental Design}.
In \bbooktitle{Advances in Neural Information Processing Systems}
\bvolume{6(9)}
\bpages{679--686}.
\bpublisher{Morgan Kaufmann Publishers}.
\end{binproceedings}
\endbibitem

\bibitem[\protect\citeauthoryear{Cressie}{1993}]{cressie2015statistics}
\begin{bbook}[author]
\bauthor{\bsnm{Cressie},~\bfnm{Noel}\binits{N.}}
(\byear{1993}).
\btitle{Statistics for spatial data}.
\bpublisher{John Wiley \& Sons}.
\end{bbook}
\endbibitem

\bibitem[\protect\citeauthoryear{Cressie and Johannesson}{2006}]{Cressie_2006}
\begin{binproceedings}[author]
\bauthor{\bsnm{Cressie},~\bfnm{N.}\binits{N.}} \AND
  \bauthor{\bsnm{Johannesson},~\bfnm{G.}\binits{G.}}
(\byear{2006}).
\btitle{Spatial prediction for massive data sets}.
In \bbooktitle{Mastering the Data Explosion in the Earth and Environmental
  Sciences: Proceedings of the Australian Academy of Science Elizabeth and
  Frederick White Conference}
\bpages{1--11}.
\bpublisher{Australian Academy of Science}, \baddress{Canberra, Australia}.
\end{binproceedings}
\endbibitem

\bibitem[\protect\citeauthoryear{Cressie and
  Johannesson}{2008}]{cressie2008fixed}
\begin{barticle}[author]
\bauthor{\bsnm{Cressie},~\bfnm{Noel}\binits{N.}} \AND
  \bauthor{\bsnm{Johannesson},~\bfnm{Gardar}\binits{G.}}
(\byear{2008}).
\btitle{Fixed rank kriging for very large spatial data sets}.
\bjournal{Journal of the Royal Statistical Society: Series B (Statistical
  Methodology)}
\bvolume{70}
\bpages{209--226}.
\end{barticle}
\endbibitem

\bibitem[\protect\citeauthoryear{Cressie and
  Wikle}{2015}]{cressiewikle2015statistics}
\begin{bbook}[author]
\bauthor{\bsnm{Cressie},~\bfnm{Noel}\binits{N.}} \AND
  \bauthor{\bsnm{Wikle},~\bfnm{Christopher~K}\binits{C.~K.}}
(\byear{2015}).
\btitle{Statistics for spatio-temporal data}.
\bpublisher{John Wiley \& Sons}.
\end{bbook}
\endbibitem

\bibitem[\protect\citeauthoryear{Dahlhaus and
  K{\"u}nsch}{1987}]{dahlhaus1987edge}
\begin{barticle}[author]
\bauthor{\bsnm{Dahlhaus},~\bfnm{R}\binits{R.}} \AND
  \bauthor{\bsnm{K{\"u}nsch},~\bfnm{H}\binits{H.}}
(\byear{1987}).
\btitle{Edge effects and efficient parameter estimation for stationary random
  fields}.
\bjournal{Biometrika}
\bvolume{74}
\bpages{877--882}.
\end{barticle}
\endbibitem

\bibitem[\protect\citeauthoryear{Datta et~al.}{2016a}]{datta2016hierarchical}
\begin{barticle}[author]
\bauthor{\bsnm{Datta},~\bfnm{Abhirup}\binits{A.}},
  \bauthor{\bsnm{Banerjee},~\bfnm{Sudipto}\binits{S.}},
  \bauthor{\bsnm{Finley},~\bfnm{Andrew~O}\binits{A.~O.}} \AND
  \bauthor{\bsnm{Gelfand},~\bfnm{Alan~E}\binits{A.~E.}}
(\byear{2016}a).
\btitle{Hierarchical nearest-neighbor Gaussian process models for large
  geostatistical datasets}.
\bjournal{Journal of the American Statistical Association}
\bvolume{111}
\bpages{800--812}.
\end{barticle}
\endbibitem

\bibitem[\protect\citeauthoryear{Datta et~al.}{2016b}]{datta2016nonseparable}
\begin{barticle}[author]
\bauthor{\bsnm{Datta},~\bfnm{Abhirup}\binits{A.}},
  \bauthor{\bsnm{Banerjee},~\bfnm{Sudipto}\binits{S.}},
  \bauthor{\bsnm{Finley},~\bfnm{Andrew~O}\binits{A.~O.}},
  \bauthor{\bsnm{Hamm},~\bfnm{Nicholas~AS}\binits{N.~A.}},
  \bauthor{\bsnm{Schaap},~\bfnm{Martijn}\binits{M.}} \betal{et~al.}
(\byear{2016}b).
\btitle{Nonseparable dynamic nearest neighbor Gaussian process models for large
  spatio-temporal data with an application to particulate matter analysis}.
\bjournal{The Annals of Applied Statistics}
\bvolume{10}
\bpages{1286--1316}.
\end{barticle}
\endbibitem

\bibitem[\protect\citeauthoryear{Datta et~al.}{2016c}]{datta2016cholesky}
\begin{barticle}[author]
\bauthor{\bsnm{Datta},~\bfnm{Abhirup}\binits{A.}},
  \bauthor{\bsnm{Banerjee},~\bfnm{Sudipto}\binits{S.}},
  \bauthor{\bsnm{Finley},~\bfnm{Andrew~O.}\binits{A.~O.}} \AND
  \bauthor{\bsnm{Gelfand},~\bfnm{Alan~E.}\binits{A.~E.}}
(\byear{2016}c).
\btitle{On nearest-neighbor Gaussian process models for massive spatial data}.
\bjournal{Wiley Interdisciplinary Reviews: Computational Statistics}
\bvolume{8}
\bpages{162--171}.
\bdoi{10.1002/wics.1383}
\end{barticle}
\endbibitem

\bibitem[\protect\citeauthoryear{Du, Zhang and Mandrekar}{2009}]{Du:etal:09}
\begin{barticle}[author]
\bauthor{\bsnm{Du},~\bfnm{Juan}\binits{J.}},
  \bauthor{\bsnm{Zhang},~\bfnm{Hao}\binits{H.}} \AND
  \bauthor{\bsnm{Mandrekar},~\bfnm{V.~S.}\binits{V.~S.}}
(\byear{2009}).
\btitle{Fixed-domain asymptotic properties of tapered maximum likelihood
  estimators}.
\bjournal{Ann. Statist.}
\bvolume{37}
\bpages{3330--3361}.
\bdoi{10.1214/08-AOS676}
\end{barticle}
\endbibitem

\bibitem[\protect\citeauthoryear{Eidsvik et~al.}{2014}]{eidsvik2014estimation}
\begin{barticle}[author]
\bauthor{\bsnm{Eidsvik},~\bfnm{Jo}\binits{J.}},
  \bauthor{\bsnm{Shaby},~\bfnm{Benjamin~A}\binits{B.~A.}},
  \bauthor{\bsnm{Reich},~\bfnm{Brian~J}\binits{B.~J.}},
  \bauthor{\bsnm{Wheeler},~\bfnm{Matthew}\binits{M.}} \AND
  \bauthor{\bsnm{Niemi},~\bfnm{Jarad}\binits{J.}}
(\byear{2014}).
\btitle{Estimation and prediction in spatial models with block composite
  likelihoods}.
\bjournal{Journal of Computational and Graphical Statistics}
\bvolume{23}
\bpages{295--315}.
\end{barticle}
\endbibitem

\bibitem[\protect\citeauthoryear{Emery}{2009}]{emery:2009}
\begin{barticle}[author]
\bauthor{\bsnm{Emery},~\bfnm{Xavier}\binits{X.}}
(\byear{2009}).
\btitle{The kriging update equations and their application to the selection of
  neighboring data}.
\bjournal{Computational Geosciences}
\bvolume{13}
\bpages{269--280}.
\end{barticle}
\endbibitem

\bibitem[\protect\citeauthoryear{Finley, Datta and Banerjee}{2017}]{spnngp}
\begin{bmanual}[author]
\bauthor{\bsnm{Finley},~\bfnm{Andrew}\binits{A.}},
  \bauthor{\bsnm{Datta},~\bfnm{Abhirup}\binits{A.}} \AND
  \bauthor{\bsnm{Banerjee},~\bfnm{Sudipto}\binits{S.}}
(\byear{2017}).
\btitle{spNNGP: Spatial Regression Models for Large Datasets using Nearest
  Neighbor Gaussian Processes}
\bnote{R package version 0.1.1}.
\end{bmanual}
\endbibitem

\bibitem[\protect\citeauthoryear{Finley et~al.}{2009}]{finley2009improving}
\begin{barticle}[author]
\bauthor{\bsnm{Finley},~\bfnm{Andrew~O}\binits{A.~O.}},
  \bauthor{\bsnm{Sang},~\bfnm{Huiyan}\binits{H.}},
  \bauthor{\bsnm{Banerjee},~\bfnm{Sudipto}\binits{S.}} \AND
  \bauthor{\bsnm{Gelfand},~\bfnm{Alan~E}\binits{A.~E.}}
(\byear{2009}).
\btitle{Improving the performance of predictive process modeling for large
  datasets}.
\bjournal{Computational statistics \& data analysis}
\bvolume{53}
\bpages{2873--2884}.
\end{barticle}
\endbibitem

\bibitem[\protect\citeauthoryear{Finley et~al.}{2017}]{finley2017algorithm}
\begin{barticle}[author]
\bauthor{\bsnm{Finley},~\bfnm{Andrew~O.}\binits{A.~O.}},
  \bauthor{\bsnm{Datta},~\bfnm{Abhirup}\binits{A.}},
  \bauthor{\bsnm{Cook},~\bfnm{Bruce~C.}\binits{B.~C.}},
  \bauthor{\bsnm{Morton},~\bfnm{Douglas~C.}\binits{D.~C.}},
  \bauthor{\bsnm{Andersen},~\bfnm{Hans~E.}\binits{H.~E.}} \AND
  \bauthor{\bsnm{Banerjee},~\bfnm{Sudipto}\binits{S.}}
(\byear{2017}).
\btitle{Applying Nearest Neighbor Gaussian Processes to Massive Spatial Data
  Sets: Forest Canopy Height Prediction Across Tanana Valley Alaska}.
\arxiv{https://arxiv.org/pdf/1702.00434.pdf}
\end{barticle}
\endbibitem

\bibitem[\protect\citeauthoryear{Fuentes}{2007}]{fuentes2007approximate}
\begin{barticle}[author]
\bauthor{\bsnm{Fuentes},~\bfnm{Montserrat}\binits{M.}}
(\byear{2007}).
\btitle{Approximate likelihood for large irregularly spaced spatial data}.
\bjournal{Journal of the American Statistical Association}
\bvolume{102}
\bpages{321--331}.
\end{barticle}
\endbibitem

\bibitem[\protect\citeauthoryear{Furrer}{2016}]{spam}
\begin{bmanual}[author]
\bauthor{\bsnm{Furrer},~\bfnm{Reinhard}\binits{R.}}
(\byear{2016}).
\btitle{spam: SPArse Matrix}
\bnote{R package version 1.4-0}.
\end{bmanual}
\endbibitem

\bibitem[\protect\citeauthoryear{Furrer, Bachoc and Du}{2016}]{Furr:Bach:Du:16}
\begin{barticle}[author]
\bauthor{\bsnm{Furrer},~\bfnm{Reinhard}\binits{R.}},
  \bauthor{\bsnm{Bachoc},~\bfnm{François}\binits{F.}} \AND
  \bauthor{\bsnm{Du},~\bfnm{Juan}\binits{J.}}
(\byear{2016}).
\btitle{Asymptotic Properties of Multivariate Tapering for Estimation and
  Prediction}.
\bjournal{J. Multivariate Anal.}
\bvolume{149}
\bpages{177--191}.
\bdoi{10.1016/j.jmva.2016.04.006}
\end{barticle}
\endbibitem

\bibitem[\protect\citeauthoryear{Furrer, Genton and
  Nychka}{2006}]{furrer2006covariance}
\begin{barticle}[author]
\bauthor{\bsnm{Furrer},~\bfnm{Reinhard}\binits{R.}},
  \bauthor{\bsnm{Genton},~\bfnm{Marc~G}\binits{M.~G.}} \AND
  \bauthor{\bsnm{Nychka},~\bfnm{Douglas}\binits{D.}}
(\byear{2006}).
\btitle{Covariance tapering for interpolation of large spatial datasets}.
\bjournal{Journal of Computational and Graphical Statistics}
\bvolume{15}
\bpages{502--523}.
\end{barticle}
\endbibitem

\bibitem[\protect\citeauthoryear{Furrer and Sain}{2010}]{Furr:Sain:10}
\begin{barticle}[author]
\bauthor{\bsnm{Furrer},~\bfnm{R.}\binits{R.}} \AND
  \bauthor{\bsnm{Sain},~\bfnm{S.~R.}\binits{S.~R.}}
(\byear{2010}).
\btitle{{spam}: {A} Sparse Matrix {R} Package with Emphasis on {MCMC} Methods
  for {G}aussian {M}arkov Random Fields}.
\bjournal{J. Stat. Softw.}
\bvolume{36}
\bpages{1--25}.
\end{barticle}
\endbibitem

\bibitem[\protect\citeauthoryear{Gerber}{2017}]{gapfill}
\begin{bmanual}[author]
\bauthor{\bsnm{Gerber},~\bfnm{Florian}\binits{F.}}
(\byear{2017}).
\btitle{{gapfill}: Fill Missing Values in Satellite Data}
\bnote{R~package version~0.9.5}.
\end{bmanual}
\endbibitem

\bibitem[\protect\citeauthoryear{Gerber et~al.}{2018}]{Gerb:etal:16}
\begin{barticle}[author]
\bauthor{\bsnm{Gerber},~\bfnm{F.}\binits{F.}},
  \bauthor{\bsnm{Furrer},~\bfnm{R.}\binits{R.}},
  \bauthor{\bsnm{Schaepman-Strub},~\bfnm{G.}\binits{G.}},
  \bauthor{\bparticle{de} \bsnm{Jong},~\bfnm{R.}\binits{R.}} \AND
  \bauthor{\bsnm{Schaepman},~\bfnm{M.~E.}\binits{M.~E.}}
(\byear{2018}).
\btitle{{Predicting missing values in spatio-temporal satellite data}}.
\bjournal{IEEE Transactions on Geoscience and Remote Sensing}
\bvolume{56}
\bpages{2841-2853}.
\end{barticle}
\endbibitem

\bibitem[\protect\citeauthoryear{Gneiting and
  Katzfuss}{2014}]{gneiting2014probabilistic}
\begin{barticle}[author]
\bauthor{\bsnm{Gneiting},~\bfnm{Tilmann}\binits{T.}} \AND
  \bauthor{\bsnm{Katzfuss},~\bfnm{Matthias}\binits{M.}}
(\byear{2014}).
\btitle{Probabilistic forecasting}.
\bjournal{Annual Review of Statistics and Its Application}
\bvolume{1}
\bpages{125--151}.
\end{barticle}
\endbibitem

\bibitem[\protect\citeauthoryear{Gneiting and
  Raftery}{2007}]{gneiting2007strictly}
\begin{barticle}[author]
\bauthor{\bsnm{Gneiting},~\bfnm{Tilmann}\binits{T.}} \AND
  \bauthor{\bsnm{Raftery},~\bfnm{Adrian~E}\binits{A.~E.}}
(\byear{2007}).
\btitle{Strictly proper scoring rules, prediction, and estimation}.
\bjournal{Journal of the American Statistical Association}
\bvolume{102}
\bpages{359--378}.
\end{barticle}
\endbibitem

\bibitem[\protect\citeauthoryear{Gramacy}{2016}]{gramacy2016laGP}
\begin{barticle}[author]
\bauthor{\bsnm{Gramacy},~\bfnm{Robert~B.}\binits{R.~B.}}
(\byear{2016}).
\btitle{{laGP}: Large-Scale Spatial Modeling via Local Approximate Gaussian
  Processes in {R}}.
\bjournal{Journal of Statistical Software}
\bvolume{72}
\bpages{1--46}.
\bdoi{10.18637/jss.v072.i01}
\end{barticle}
\endbibitem

\bibitem[\protect\citeauthoryear{Gramacy and Apley}{2015}]{gramacy2015local}
\begin{barticle}[author]
\bauthor{\bsnm{Gramacy},~\bfnm{R.~B.}\binits{R.~B.}} \AND
  \bauthor{\bsnm{Apley},~\bfnm{D.}\binits{D.}}
(\byear{2015}).
\btitle{Local Gaussian Process Approximation for Large Computer Experiments}.
\bjournal{Journal of Computational and Graphical Statistics}
\bvolume{24}
\bpages{561--578}.
\end{barticle}
\endbibitem

\bibitem[\protect\citeauthoryear{Gramacy and
  Haaland}{2016}]{gramacy2015speeding}
\begin{barticle}[author]
\bauthor{\bsnm{Gramacy},~\bfnm{Robert~B}\binits{R.~B.}} \AND
  \bauthor{\bsnm{Haaland},~\bfnm{Benjamin}\binits{B.}}
(\byear{2016}).
\btitle{Speeding up neighborhood search in local Gaussian process prediction}.
\bjournal{Technometrics}
\bvolume{58}
\bpages{294--303}.
\end{barticle}
\endbibitem

\bibitem[\protect\citeauthoryear{Gramacy, Niemi and
  Weiss}{2014}]{gramacy:niemi:weiss:2014}
\begin{barticle}[author]
\bauthor{\bsnm{Gramacy},~\bfnm{R.~B.}\binits{R.~B.}},
  \bauthor{\bsnm{Niemi},~\bfnm{J.}\binits{J.}} \AND
  \bauthor{\bsnm{Weiss},~\bfnm{R.}\binits{R.}}
(\byear{2014}).
\btitle{Massively Parallel Approximate Gaussian Process Regression}.
\bjournal{Journal of Uncertainty Quantification}
\bvolume{2}
\bpages{564--584}.
\end{barticle}
\endbibitem

\bibitem[\protect\citeauthoryear{Guhaniyogi and
  Banerjee}{2018}]{guhaniyogi2018meta}
\begin{barticle}[author]
\bauthor{\bsnm{Guhaniyogi},~\bfnm{Rajarshi}\binits{R.}} \AND
  \bauthor{\bsnm{Banerjee},~\bfnm{Sudipto}\binits{S.}}
(\byear{2018}).
\btitle{Meta-kriging: Scalable Bayesian modeling and inference for massive
  spatial datasets}.
\bjournal{Technometrics}
\bvolume{just-accepted}.
\end{barticle}
\endbibitem

\bibitem[\protect\citeauthoryear{Guhaniyogi
  et~al.}{2017}]{guhaniyogi2017divide}
\begin{barticle}[author]
\bauthor{\bsnm{Guhaniyogi},~\bfnm{Rajarshi}\binits{R.}},
  \bauthor{\bsnm{Li},~\bfnm{Cheng}\binits{C.}},
  \bauthor{\bsnm{Savitsky},~\bfnm{Terrance~D}\binits{T.~D.}} \AND
  \bauthor{\bsnm{Srivastava},~\bfnm{Sanvesh}\binits{S.}}
(\byear{2017}).
\btitle{A Divide-and-Conquer Bayesian Approach to Large-Scale Kriging}.
\bjournal{arXiv preprint arXiv:1712.09767}.
\end{barticle}
\endbibitem

\bibitem[\protect\citeauthoryear{Guinness}{2017}]{guinness2017spectral}
\begin{barticle}[author]
\bauthor{\bsnm{Guinness},~\bfnm{Joseph}\binits{J.}}
(\byear{2017}).
\btitle{Spectral Density Estimation for Random Fields via Periodic Embeddings}.
\bjournal{arXiv preprint arXiv:1710.08978}.
\end{barticle}
\endbibitem

\bibitem[\protect\citeauthoryear{Guinness and
  Fuentes}{2017}]{guinness2017circulant}
\begin{barticle}[author]
\bauthor{\bsnm{Guinness},~\bfnm{Joseph}\binits{J.}} \AND
  \bauthor{\bsnm{Fuentes},~\bfnm{Montserrat}\binits{M.}}
(\byear{2017}).
\btitle{Circulant embedding of approximate covariances for inference from
  {G}aussian data on large lattices}.
\bjournal{Journal of Computational and Graphical Statistics}
\bvolume{26}
\bpages{88--97}.
\end{barticle}
\endbibitem

\bibitem[\protect\citeauthoryear{Guyon}{1982}]{guyon1982parameter}
\begin{barticle}[author]
\bauthor{\bsnm{Guyon},~\bfnm{Xavier}\binits{X.}}
(\byear{1982}).
\btitle{Parameter estimation for a stationary process on a d-dimensional
  lattice}.
\bjournal{Biometrika}
\bvolume{69}
\bpages{95--105}.
\end{barticle}
\endbibitem

\bibitem[\protect\citeauthoryear{Heaton, Christensen and
  Terres}{2017}]{heaton2017nonstationary}
\begin{barticle}[author]
\bauthor{\bsnm{Heaton},~\bfnm{Matthew~J}\binits{M.~J.}},
  \bauthor{\bsnm{Christensen},~\bfnm{William~F}\binits{W.~F.}} \AND
  \bauthor{\bsnm{Terres},~\bfnm{Maria~A}\binits{M.~A.}}
(\byear{2017}).
\btitle{Nonstationary Gaussian process models using spatial hierarchical
  clustering from finite differences}.
\bjournal{Technometrics}
\bvolume{59}
\bpages{93--101}.
\end{barticle}
\endbibitem

\bibitem[\protect\citeauthoryear{Higdon}{2002}]{higdon2002space}
\begin{bincollection}[author]
\bauthor{\bsnm{Higdon},~\bfnm{Dave}\binits{D.}}
(\byear{2002}).
\btitle{Space and space-time modeling using process convolutions}.
In \bbooktitle{Quantitative methods for current environmental issues}
\bpages{37--56}.
\bpublisher{Springer}.
\end{bincollection}
\endbibitem

\bibitem[\protect\citeauthoryear{Hirano and Yajima}{2013}]{Hira:Yaji:13}
\begin{barticle}[author]
\bauthor{\bsnm{Hirano},~\bfnm{Toshihiro}\binits{T.}} \AND
  \bauthor{\bsnm{Yajima},~\bfnm{Yoshihiro}\binits{Y.}}
(\byear{2013}).
\btitle{Covariance tapering for prediction of large spatial data sets in
  transformed random fields}.
\bjournal{Annals of the Institute of Statistical Mathematics}
\bvolume{65}
\bpages{913--939}.
\bdoi{10.1007/s10463-013-0399-8}
\end{barticle}
\endbibitem

\bibitem[\protect\citeauthoryear{Kang and Cressie}{2011}]{kang2011bayesian}
\begin{barticle}[author]
\bauthor{\bsnm{Kang},~\bfnm{Emily~L}\binits{E.~L.}} \AND
  \bauthor{\bsnm{Cressie},~\bfnm{Noel}\binits{N.}}
(\byear{2011}).
\btitle{Bayesian inference for the spatial random effects model}.
\bjournal{Journal of the American Statistical Association}
\bvolume{106}
\bpages{972--983}.
\end{barticle}
\endbibitem

\bibitem[\protect\citeauthoryear{Katzfuss}{2017}]{Katzfuss2015}
\begin{barticle}[author]
\bauthor{\bsnm{Katzfuss},~\bfnm{Matthias}\binits{M.}}
(\byear{2017}).
\btitle{{A multi-resolution approximation for massive spatial datasets}}.
\bjournal{Journal of the American Statistical Association}
\bvolume{112}
\bpages{201--214}.
\bdoi{10.1080/01621459.2015.1123632}
\end{barticle}
\endbibitem

\bibitem[\protect\citeauthoryear{Katzfuss and
  Cressie}{2011}]{katzfuss2011spatio}
\begin{barticle}[author]
\bauthor{\bsnm{Katzfuss},~\bfnm{Matthias}\binits{M.}} \AND
  \bauthor{\bsnm{Cressie},~\bfnm{Noel}\binits{N.}}
(\byear{2011}).
\btitle{Spatio-temporal smoothing and EM estimation for massive remote-sensing
  data sets}.
\bjournal{Journal of Time Series Analysis}
\bvolume{32}
\bpages{430--446}.
\end{barticle}
\endbibitem

\bibitem[\protect\citeauthoryear{Katzfuss and Gong}{2017}]{KatzfussGong2017}
\begin{barticle}[author]
\bauthor{\bsnm{Katzfuss},~\bfnm{Matthias}\binits{M.}} \AND
  \bauthor{\bsnm{Gong},~\bfnm{Wenlong}\binits{W.}}
(\byear{2017}).
\btitle{{Multi-resolution approximations of Gaussian processes for large
  spatial datasets}}.
\bjournal{arXiv:1710.08976}.
\end{barticle}
\endbibitem

\bibitem[\protect\citeauthoryear{Katzfuss and
  Hammerling}{2017}]{katzfuss2014parallel}
\begin{barticle}[author]
\bauthor{\bsnm{Katzfuss},~\bfnm{Matthias}\binits{M.}} \AND
  \bauthor{\bsnm{Hammerling},~\bfnm{Dorit}\binits{D.}}
(\byear{2017}).
\btitle{{Parallel inference for massive distributed spatial data using low-rank
  models}}.
\bjournal{Statistics and Computing}
\bvolume{27}
\bpages{363--375}.
\end{barticle}
\endbibitem

\bibitem[\protect\citeauthoryear{Kaufman, Schervish and
  Nychka}{2008}]{kaufman2008covariance}
\begin{barticle}[author]
\bauthor{\bsnm{Kaufman},~\bfnm{Cari~G}\binits{C.~G.}},
  \bauthor{\bsnm{Schervish},~\bfnm{Mark~J}\binits{M.~J.}} \AND
  \bauthor{\bsnm{Nychka},~\bfnm{Douglas~W}\binits{D.~W.}}
(\byear{2008}).
\btitle{Covariance tapering for likelihood-based estimation in large spatial
  data sets}.
\bjournal{Journal of the American Statistical Association}
\bvolume{103}
\bpages{1545--1555}.
\end{barticle}
\endbibitem

\bibitem[\protect\citeauthoryear{Kim, Mallick and
  Holmes}{2005}]{kim2005analyzing}
\begin{barticle}[author]
\bauthor{\bsnm{Kim},~\bfnm{Hyoung-Moon}\binits{H.-M.}},
  \bauthor{\bsnm{Mallick},~\bfnm{Bani~K}\binits{B.~K.}} \AND
  \bauthor{\bsnm{Holmes},~\bfnm{CC}\binits{C.}}
(\byear{2005}).
\btitle{Analyzing nonstationary spatial data using piecewise Gaussian
  processes}.
\bjournal{Journal of the American Statistical Association}
\bvolume{100}
\bpages{653--668}.
\end{barticle}
\endbibitem

\bibitem[\protect\citeauthoryear{Kleiber and
  Nychka}{2015}]{kleiber2015equivalent}
\begin{barticle}[author]
\bauthor{\bsnm{Kleiber},~\bfnm{William}\binits{W.}} \AND
  \bauthor{\bsnm{Nychka},~\bfnm{Douglas~W}\binits{D.~W.}}
(\byear{2015}).
\btitle{Equivalent kriging}.
\bjournal{Spatial Statistics}
\bvolume{12}
\bpages{31--49}.
\end{barticle}
\endbibitem

\bibitem[\protect\citeauthoryear{Knorr-Held and
  Ra{\ss}er}{2000}]{knorr2000bayesian}
\begin{barticle}[author]
\bauthor{\bsnm{Knorr-Held},~\bfnm{Leonhard}\binits{L.}} \AND
  \bauthor{\bsnm{Ra{\ss}er},~\bfnm{G{\"u}nter}\binits{G.}}
(\byear{2000}).
\btitle{Bayesian detection of clusters and discontinuities in disease maps}.
\bjournal{Biometrics}
\bvolume{56}
\bpages{13--21}.
\end{barticle}
\endbibitem

\bibitem[\protect\citeauthoryear{Konomi, Sang and
  Mallick}{2014}]{konomi2014adaptive}
\begin{barticle}[author]
\bauthor{\bsnm{Konomi},~\bfnm{Bledar~A}\binits{B.~A.}},
  \bauthor{\bsnm{Sang},~\bfnm{Huiyan}\binits{H.}} \AND
  \bauthor{\bsnm{Mallick},~\bfnm{Bani~K}\binits{B.~K.}}
(\byear{2014}).
\btitle{Adaptive bayesian nonstationary modeling for large spatial datasets
  using covariance approximations}.
\bjournal{Journal of Computational and Graphical Statistics}
\bvolume{23}
\bpages{802--829}.
\end{barticle}
\endbibitem

\bibitem[\protect\citeauthoryear{Lemos and Sans{\'o}}{2009}]{lemos2009spatio}
\begin{barticle}[author]
\bauthor{\bsnm{Lemos},~\bfnm{Ricardo~T}\binits{R.~T.}} \AND
  \bauthor{\bsnm{Sans{\'o}},~\bfnm{Bruno}\binits{B.}}
(\byear{2009}).
\btitle{A spatio-temporal model for mean, anomaly, and trend fields of North
  Atlantic sea surface temperature}.
\bjournal{Journal of the American Statistical Association}
\bvolume{104}
\bpages{5--18}.
\end{barticle}
\endbibitem

\bibitem[\protect\citeauthoryear{Liang et~al.}{2013}]{liang2013resampling}
\begin{barticle}[author]
\bauthor{\bsnm{Liang},~\bfnm{Faming}\binits{F.}},
  \bauthor{\bsnm{Cheng},~\bfnm{Yichen}\binits{Y.}},
  \bauthor{\bsnm{Song},~\bfnm{Qifan}\binits{Q.}},
  \bauthor{\bsnm{Park},~\bfnm{Jincheol}\binits{J.}} \AND
  \bauthor{\bsnm{Yang},~\bfnm{Ping}\binits{P.}}
(\byear{2013}).
\btitle{A resampling-based stochastic approximation method for analysis of
  large geostatistical data}.
\bjournal{Journal of the American Statistical Association}
\bvolume{108}
\bpages{325--339}.
\end{barticle}
\endbibitem

\bibitem[\protect\citeauthoryear{Lindgren, Rue and
  Lindstr{\"o}m}{2011}]{lindgren2011explicit}
\begin{barticle}[author]
\bauthor{\bsnm{Lindgren},~\bfnm{Finn}\binits{F.}},
  \bauthor{\bsnm{Rue},~\bfnm{H{\aa}vard}\binits{H.}} \AND
  \bauthor{\bsnm{Lindstr{\"o}m},~\bfnm{Johan}\binits{J.}}
(\byear{2011}).
\btitle{An explicit link between Gaussian fields and Gaussian Markov random
  fields: the stochastic partial differential equation approach}.
\bjournal{Journal of the Royal Statistical Society: Series B (Statistical
  Methodology)}
\bvolume{73}
\bpages{423--498}.
\end{barticle}
\endbibitem

\bibitem[\protect\citeauthoryear{Litvinenko
  et~al.}{2017}]{litvinenko2017likelihood}
\begin{barticle}[author]
\bauthor{\bsnm{Litvinenko},~\bfnm{Alexander}\binits{A.}},
  \bauthor{\bsnm{Sun},~\bfnm{Ying}\binits{Y.}},
  \bauthor{\bsnm{Genton},~\bfnm{Marc~G}\binits{M.~G.}} \AND
  \bauthor{\bsnm{Keyes},~\bfnm{David}\binits{D.}}
(\byear{2017}).
\btitle{Likelihood Approximation With Hierarchical Matrices For Large Spatial
  Datasets}.
\bjournal{arXiv preprint arXiv:1709.04419}.
\end{barticle}
\endbibitem

\bibitem[\protect\citeauthoryear{Minsker}{2015}]{minsker2015geometric}
\begin{barticle}[author]
\bauthor{\bsnm{Minsker},~\bfnm{Stanislav}\binits{S.}}
(\byear{2015}).
\btitle{Geometric median and robust estimation in Banach spaces}.
\bjournal{Bernoulli}
\bvolume{21}
\bpages{2308--2335}.
\end{barticle}
\endbibitem

\bibitem[\protect\citeauthoryear{Minsker et~al.}{2014}]{minsker2014robust}
\begin{barticle}[author]
\bauthor{\bsnm{Minsker},~\bfnm{Stanislav}\binits{S.}},
  \bauthor{\bsnm{Srivastava},~\bfnm{Sanvesh}\binits{S.}},
  \bauthor{\bsnm{Lin},~\bfnm{Lizhen}\binits{L.}} \AND
  \bauthor{\bsnm{Dunson},~\bfnm{David~B}\binits{D.~B.}}
(\byear{2014}).
\btitle{Robust and scalable Bayes via a median of subset posterior measures}.
\bjournal{arXiv preprint arXiv:1403.2660}.
\end{barticle}
\endbibitem

\bibitem[\protect\citeauthoryear{Neelon, Gelfand and
  Miranda}{2014}]{neelon2014multivariate}
\begin{barticle}[author]
\bauthor{\bsnm{Neelon},~\bfnm{Brian}\binits{B.}},
  \bauthor{\bsnm{Gelfand},~\bfnm{Alan~E}\binits{A.~E.}} \AND
  \bauthor{\bsnm{Miranda},~\bfnm{Marie~Lynn}\binits{M.~L.}}
(\byear{2014}).
\btitle{A multivariate spatial mixture model for areal data: examining regional
  differences in standardized test scores}.
\bjournal{Journal of the Royal Statistical Society: Series C (Applied
  Statistics)}
\bvolume{63}
\bpages{737--761}.
\end{barticle}
\endbibitem

\bibitem[\protect\citeauthoryear{Nychka
  et~al.}{2015}]{nychka2015multiresolution}
\begin{barticle}[author]
\bauthor{\bsnm{Nychka},~\bfnm{Douglas}\binits{D.}},
  \bauthor{\bsnm{Bandyopadhyay},~\bfnm{Soutir}\binits{S.}},
  \bauthor{\bsnm{Hammerling},~\bfnm{Dorit}\binits{D.}},
  \bauthor{\bsnm{Lindgren},~\bfnm{Finn}\binits{F.}} \AND
  \bauthor{\bsnm{Sain},~\bfnm{Stephan}\binits{S.}}
(\byear{2015}).
\btitle{A multiresolution Gaussian process model for the analysis of large
  spatial datasets}.
\bjournal{Journal of Computational and Graphical Statistics}
\bvolume{24}
\bpages{579--599}.
\end{barticle}
\endbibitem

\bibitem[\protect\citeauthoryear{Paciorek
  et~al.}{2015}]{paciorek2013parallelizing}
\begin{barticle}[author]
\bauthor{\bsnm{Paciorek},~\bfnm{Christopher~J}\binits{C.~J.}},
  \bauthor{\bsnm{Lipshitz},~\bfnm{Benjamin}\binits{B.}},
  \bauthor{\bsnm{Zhuo},~\bfnm{Wei}\binits{W.}},
  \bauthor{\bsnm{Kaufman},~\bfnm{Cari~G}\binits{C.~G.}},
  \bauthor{\bsnm{Thomas},~\bfnm{Rollin~C}\binits{R.~C.}} \betal{et~al.}
(\byear{2015}).
\btitle{Parallelizing Gaussian Process Calculations In R}.
\bjournal{Journal of Statistical Software}
\bvolume{63}
\bpages{1-23}.
\end{barticle}
\endbibitem

\bibitem[\protect\citeauthoryear{Rue, Martino and Chopin}{2009}]{rue2009inla}
\begin{barticle}[author]
\bauthor{\bsnm{Rue},~\bfnm{H{\aa}vard}\binits{H.}},
  \bauthor{\bsnm{Martino},~\bfnm{Sara}\binits{S.}} \AND
  \bauthor{\bsnm{Chopin},~\bfnm{Nicolas}\binits{N.}}
(\byear{2009}).
\btitle{Approximate {B}ayesian inference for latent {G}aussian models by using
  integrated nested {L}aplace approximations}.
\bjournal{Journal of the Royal Statistical Society: Series B (Statistical
  Methodology)}
\bvolume{71}
\bpages{319--392}.
\bdoi{10.1111/j.1467-9868.2008.00700.x}
\end{barticle}
\endbibitem

\bibitem[\protect\citeauthoryear{Rue et~al.}{2017}]{RINLA}
\begin{bmanual}[author]
\bauthor{\bsnm{Rue},~\bfnm{Havard}\binits{H.}},
  \bauthor{\bsnm{Martino},~\bfnm{Sara}\binits{S.}},
  \bauthor{\bsnm{Lindgren},~\bfnm{Finn}\binits{F.}},
  \bauthor{\bsnm{Simpson},~\bfnm{Daniel}\binits{D.}},
  \bauthor{\bsnm{Riebler},~\bfnm{Andrea}\binits{A.}},
  \bauthor{\bsnm{Krainski},~\bfnm{Elias~Teixeira}\binits{E.~T.}} \AND
  \bauthor{\bsnm{Fuglstad},~\bfnm{Geir-Arne}\binits{G.-A.}}
(\byear{2017}).
\btitle{INLA: Bayesian Analysis of Latent Gaussian Models using Integrated
  Nested Laplace Approximations}
\bnote{R package version 17.06.20}.
\end{bmanual}
\endbibitem

\bibitem[\protect\citeauthoryear{Sang and Huang}{2012}]{sang2012full}
\begin{barticle}[author]
\bauthor{\bsnm{Sang},~\bfnm{Huiyan}\binits{H.}} \AND
  \bauthor{\bsnm{Huang},~\bfnm{Jianhua~Z}\binits{J.~Z.}}
(\byear{2012}).
\btitle{A full scale approximation of covariance functions for large spatial
  data sets}.
\bjournal{Journal of the Royal Statistical Society: Series B (Statistical
  Methodology)}
\bvolume{74}
\bpages{111--132}.
\end{barticle}
\endbibitem

\bibitem[\protect\citeauthoryear{Sang, Jun and
  Huang}{2011}]{sang2011covariance}
\begin{barticle}[author]
\bauthor{\bsnm{Sang},~\bfnm{Huiyan}\binits{H.}},
  \bauthor{\bsnm{Jun},~\bfnm{Mikyoung}\binits{M.}} \AND
  \bauthor{\bsnm{Huang},~\bfnm{Jianhua~Z}\binits{J.~Z.}}
(\byear{2011}).
\btitle{Covariance approximation for large multivariate spatial data sets with
  an application to multiple climate model errors}.
\bjournal{The Annals of Applied Statistics}
\bpages{2519--2548}.
\end{barticle}
\endbibitem

\bibitem[\protect\citeauthoryear{Schabenberger and
  Gotway}{2004}]{schabenberger2004statistical}
\begin{bbook}[author]
\bauthor{\bsnm{Schabenberger},~\bfnm{Oliver}\binits{O.}} \AND
  \bauthor{\bsnm{Gotway},~\bfnm{Carol~A}\binits{C.~A.}}
(\byear{2004}).
\btitle{Statistical methods for spatial data analysis}.
\bpublisher{CRC press}.
\end{bbook}
\endbibitem

\bibitem[\protect\citeauthoryear{Simpson, Lindgren and
  Rue}{2012}]{simpson2012order}
\begin{barticle}[author]
\bauthor{\bsnm{Simpson},~\bfnm{Daniel}\binits{D.}},
  \bauthor{\bsnm{Lindgren},~\bfnm{Finn}\binits{F.}} \AND
  \bauthor{\bsnm{Rue},~\bfnm{H{\aa}vard}\binits{H.}}
(\byear{2012}).
\btitle{In order to make spatial statistics computationally feasible, we need
  to forget about the covariance function}.
\bjournal{Environmetrics}
\bvolume{23}
\bpages{65--74}.
\end{barticle}
\endbibitem

\bibitem[\protect\citeauthoryear{Stein}{1999}]{Stei:99}
\begin{bbook}[author]
\bauthor{\bsnm{Stein},~\bfnm{M.~L.}\binits{M.~L.}}
(\byear{1999}).
\btitle{Interpolation of Spatial Data}.
\bpublisher{Springer-Verlag}
\bnote{Some theory for Kriging}.
\end{bbook}
\endbibitem

\bibitem[\protect\citeauthoryear{Stein}{2013}]{stein2013statistical}
\begin{barticle}[author]
\bauthor{\bsnm{Stein},~\bfnm{Michael~L}\binits{M.~L.}}
(\byear{2013}).
\btitle{Statistical properties of covariance tapers}.
\bjournal{Journal of Computational and Graphical Statistics}
\bvolume{22}
\bpages{866--885}.
\end{barticle}
\endbibitem

\bibitem[\protect\citeauthoryear{Stein}{2014}]{stein2014limitations}
\begin{barticle}[author]
\bauthor{\bsnm{Stein},~\bfnm{Michael~L}\binits{M.~L.}}
(\byear{2014}).
\btitle{Limitations on low rank approximations for covariance matrices of
  spatial data}.
\bjournal{Spatial Statistics}
\bvolume{8}
\bpages{1--19}.
\end{barticle}
\endbibitem

\bibitem[\protect\citeauthoryear{Stein, Chi and
  Welty}{2004}]{stein2004approximating}
\begin{barticle}[author]
\bauthor{\bsnm{Stein},~\bfnm{Michael~L}\binits{M.~L.}},
  \bauthor{\bsnm{Chi},~\bfnm{Zhiyi}\binits{Z.}} \AND
  \bauthor{\bsnm{Welty},~\bfnm{Leah~J}\binits{L.~J.}}
(\byear{2004}).
\btitle{Approximating likelihoods for large spatial data sets}.
\bjournal{Journal of the Royal Statistical Society: Series B (Statistical
  Methodology)}
\bvolume{66}
\bpages{275--296}.
\end{barticle}
\endbibitem

\bibitem[\protect\citeauthoryear{Stein et~al.}{2013}]{stein2013stochastic}
\begin{barticle}[author]
\bauthor{\bsnm{Stein},~\bfnm{Michael~L}\binits{M.~L.}},
  \bauthor{\bsnm{Chen},~\bfnm{Jie}\binits{J.}},
  \bauthor{\bsnm{Anitescu},~\bfnm{Mihai}\binits{M.}} \betal{et~al.}
(\byear{2013}).
\btitle{Stochastic approximation of score functions for Gaussian processes}.
\bjournal{The Annals of Applied Statistics}
\bvolume{7}
\bpages{1162--1191}.
\end{barticle}
\endbibitem

\bibitem[\protect\citeauthoryear{Sun, Li and
  Genton}{2012}]{sun2012geostatistics}
\begin{bincollection}[author]
\bauthor{\bsnm{Sun},~\bfnm{Ying}\binits{Y.}},
  \bauthor{\bsnm{Li},~\bfnm{Bo}\binits{B.}} \AND
  \bauthor{\bsnm{Genton},~\bfnm{Marc~G}\binits{M.~G.}}
(\byear{2012}).
\btitle{Geostatistics for large datasets}.
In \bbooktitle{Advances and challenges in space-time modelling of natural
  events}
\bpages{55--77}.
\bpublisher{Springer}.
\end{bincollection}
\endbibitem

\bibitem[\protect\citeauthoryear{Sun and Stein}{2016}]{sun2016statistically}
\begin{barticle}[author]
\bauthor{\bsnm{Sun},~\bfnm{Ying}\binits{Y.}} \AND
  \bauthor{\bsnm{Stein},~\bfnm{Michael~L}\binits{M.~L.}}
(\byear{2016}).
\btitle{Statistically and computationally efficient estimating equations for
  large spatial datasets}.
\bjournal{Journal of Computational and Graphical Statistics}
\bvolume{25}
\bpages{187--208}.
\end{barticle}
\endbibitem

\bibitem[\protect\citeauthoryear{Taylor-Rodriguez
  et~al.}{2018}]{taylor2018spatial}
\begin{barticle}[author]
\bauthor{\bsnm{Taylor-Rodriguez},~\bfnm{Daniel}\binits{D.}},
  \bauthor{\bsnm{Finley},~\bfnm{Andrew~O}\binits{A.~O.}},
  \bauthor{\bsnm{Datta},~\bfnm{Abhirup}\binits{A.}},
  \bauthor{\bsnm{Babcock},~\bfnm{Chad}\binits{C.}},
  \bauthor{\bsnm{Andersen},~\bfnm{Hans-Erik}\binits{H.-E.}},
  \bauthor{\bsnm{Cook},~\bfnm{Bruce~D}\binits{B.~D.}},
  \bauthor{\bsnm{Morton},~\bfnm{Douglas~C}\binits{D.~C.}} \AND
  \bauthor{\bsnm{Baneerjee},~\bfnm{Sudipto}\binits{S.}}
(\byear{2018}).
\btitle{Spatial Factor Models for High-Dimensional and Large Spatial Data: An
  Application in Forest Variable Mapping}.
\bjournal{arXiv preprint arXiv:1801.02078}.
\end{barticle}
\endbibitem

\bibitem[\protect\citeauthoryear{Ton et~al.}{2017}]{ton2017spatial}
\begin{barticle}[author]
\bauthor{\bsnm{Ton},~\bfnm{Jean-Francois}\binits{J.-F.}},
  \bauthor{\bsnm{Flaxman},~\bfnm{Seth}\binits{S.}},
  \bauthor{\bsnm{Sejdinovic},~\bfnm{Dino}\binits{D.}} \AND
  \bauthor{\bsnm{Bhatt},~\bfnm{Samir}\binits{S.}}
(\byear{2017}).
\btitle{Spatial Mapping with Gaussian Processes and Nonstationary Fourier
  Features}.
\bjournal{arXiv preprint arXiv:1711.05615}.
\end{barticle}
\endbibitem

\bibitem[\protect\citeauthoryear{Vapnik}{1995}]{vapnik:1995}
\begin{bbook}[author]
\bauthor{\bsnm{Vapnik},~\bfnm{V.~N.}\binits{V.~N.}}
(\byear{1995}).
\btitle{The Nature of Statistical Learning Theory}.
\bpublisher{Springer Verlag}, \baddress{New York}.
\end{bbook}
\endbibitem

\bibitem[\protect\citeauthoryear{Varin, Reid and
  Firth}{2011}]{varin2011overview}
\begin{barticle}[author]
\bauthor{\bsnm{Varin},~\bfnm{Cristiano}\binits{C.}},
  \bauthor{\bsnm{Reid},~\bfnm{Nancy}\binits{N.}} \AND
  \bauthor{\bsnm{Firth},~\bfnm{David}\binits{D.}}
(\byear{2011}).
\btitle{An overview of composite likelihood methods}.
\bjournal{Statistica Sinica}
\bpages{5--42}.
\end{barticle}
\endbibitem

\bibitem[\protect\citeauthoryear{Vecchia}{1988}]{vecchia1988estimation}
\begin{barticle}[author]
\bauthor{\bsnm{Vecchia},~\bfnm{Aldo~V}\binits{A.~V.}}
(\byear{1988}).
\btitle{Estimation and model identification for continuous spatial processes}.
\bjournal{Journal of the Royal Statistical Society. Series B (Methodological)}
\bpages{297--312}.
\end{barticle}
\endbibitem

\bibitem[\protect\citeauthoryear{Wang and Loh}{2011}]{Wang:Loh:11}
\begin{barticle}[author]
\bauthor{\bsnm{Wang},~\bfnm{Daqing}\binits{D.}} \AND
  \bauthor{\bsnm{Loh},~\bfnm{Wei-Liem}\binits{W.-L.}}
(\byear{2011}).
\btitle{On fixed-domain asymptotics and covariance tapering in Gaussian random
  field models}.
\bjournal{Electron. J. Statist.}
\bvolume{5}
\bpages{238--269}.
\bdoi{10.1214/11-EJS607}
\end{barticle}
\endbibitem

\bibitem[\protect\citeauthoryear{Weiss et~al.}{2014}]{Weis:etal:14}
\begin{barticle}[author]
\bauthor{\bsnm{Weiss},~\bfnm{Daniel~J.}\binits{D.~J.}},
  \bauthor{\bsnm{Atkinson},~\bfnm{Peter~M.}\binits{P.~M.}},
  \bauthor{\bsnm{Bhatt},~\bfnm{Samir}\binits{S.}},
  \bauthor{\bsnm{Mappin},~\bfnm{Bonnie}\binits{B.}},
  \bauthor{\bsnm{Hay},~\bfnm{Simon~I.}\binits{S.~I.}} \AND
  \bauthor{\bsnm{Gething},~\bfnm{Peter~W.}\binits{P.~W.}}
(\byear{2014}).
\btitle{An effective approach for gap-filling continental scale remotely sensed
  time-series}.
\bjournal{{ISPRS} J. Photogramm. Remote Sens.}
\bvolume{98}
\bpages{106--118}.
\bdoi{10.1016/j.isprsjprs.2014.10.001}
\end{barticle}
\endbibitem

\bibitem[\protect\citeauthoryear{Whittle}{1954}]{whittle1954stationary}
\begin{barticle}[author]
\bauthor{\bsnm{Whittle},~\bfnm{Peter}\binits{P.}}
(\byear{1954}).
\btitle{On stationary processes in the plane}.
\bjournal{Biometrika}
\bpages{434--449}.
\end{barticle}
\endbibitem

\bibitem[\protect\citeauthoryear{Wikle et~al.}{2017}]{wikle2017common}
\begin{barticle}[author]
\bauthor{\bsnm{Wikle},~\bfnm{Christopher~K}\binits{C.~K.}},
  \bauthor{\bsnm{Cressie},~\bfnm{Noel}\binits{N.}},
  \bauthor{\bsnm{Zammit-Mangion},~\bfnm{Andrew}\binits{A.}} \AND
  \bauthor{\bsnm{Shumack},~\bfnm{Clint}\binits{C.}}
(\byear{2017}).
\btitle{A Common Task Framework (CTF) for Objective Comparison of Spatial
  Prediction Methodologies}.
\bjournal{Statistics Views}.
\end{barticle}
\endbibitem

\bibitem[\protect\citeauthoryear{Zammit-Mangion and
  Cressie}{2017}]{zammit2017frk}
\begin{barticle}[author]
\bauthor{\bsnm{Zammit-Mangion},~\bfnm{Andrew}\binits{A.}} \AND
  \bauthor{\bsnm{Cressie},~\bfnm{Noel}\binits{N.}}
(\byear{2017}).
\btitle{FRK: An R Package for Spatial and Spatio-Temporal Prediction with Large
  Datasets}.
\bjournal{arXiv preprint arXiv:1705.08105}.
\end{barticle}
\endbibitem

\end{thebibliography}

\end{document}